\documentclass[10pt,onecolumn,
preprint
]{aastex}
\usepackage{url}

\usepackage{latexsym}
\usepackage{bm}
\usepackage{calrsfs}
\usepackage{graphicx}

\RequirePackage{color}

\newcommand{\emaila}{namouni@obs-nice.fr}

\begin{document}

\title{On dynamical friction in a gaseous medium with a boundary}

%%

%\slugcomment{Astrophysics and Space Science, in press}

%% Running heads
\shorttitle{Dynamical friction in a gaseous medium}
\shortauthors{Namouni}

\author{Fathi Namouni} 
\affil{Universit\'e de Nice, CNRS,  Observatoire de la C\^ote d'Azur, BP 4229, 06304 Nice, France}
\email{\emaila}

\begin{abstract}
Dynamical friction arises from the interaction of a perturber and the gravitational wake it excites in the ambient medium. We study the effects of the presence of a boundary on dynamical friction by studying analytically the interaction of  perturber with uniform rectilinear motion in a uniform homogeneous medium with a reflecting planar boundary. Wake reflection at a medium's boundary may occur at the edges of truncated disks perturbed by planetary or stellar companions as well as in numerical simulations of planet-disk interaction with no-outflow boundary conditions. In this paper, we show that the presence of the boundary modifies the behaviour of dynamical friction significantly. We find that perturbers are invariably pushed away from the boundary and reach a terminal subsonic velocity near Mach 0.37 regardless of  initial velocity. Dynamical friction may even be reversed for Mach numbers less than 0.37 thereby accelerating instead of decelerating the perturber. Perturbers moving  parallel to the boundary feel additional friction orthogonal to the direction of motion that is much stronger than the standard friction along the direction of motion. These results indicate that the common use of the standard Chandrasekhar formula as a short hand estimate of dynamical friction may be inadequate as observed in various numerical simulations.
\end{abstract}

\newcommand{\rmx}{r_{\rm max}}

\section{Introduction}

Dynamical friction, the reaction force a perturber feels from the gravitational wake it excites in the ambient medium,   is one of the fundamental processes in the formation of astronomical structures. Applications of dynamical friction to collisionless systems include mass segregation in star clusters \citep{r1,r2}, galaxy distribution in hierarchical cluster formation \citep{r3,r4}, the decay of satellites in galactic halos and galaxy mergers \citep{r5} (and references therein).  Dynamical friction also occurs in collisional particle systems and was applied to the study of the growth of planetesimals \citep{r6,r7}, the eccentricity excitation of planetary embryo orbits \citep{r8,r9,r10}, and the confinement of planetary rings \citep{r11}. For gaseous systems, applications include the orbital decay of compact stars orbiting supermassive black holes \citep{r12}, the heating of intercluster gas by decelerating supersonic galaxies \citep{r13,r14,r15}, the characterization of the differences in X-ray emissions in galaxy clusters between the standard cold dark matter model and modified newtonian dynamics \citep{r16}.  

Boundaries appear in dynamical friction because both the perturber and the ambient medium that surrounds it have finite sizes. In the common analyses of dynamical friction, these two boundaries enter the force through the Coulomb logarithm that requires cutoffs at small and large distances from the perturber to avoid divergence. In the context of collisionless systems,  the classical derivation of dynamical friction using a perturber in uniform rectilinear motion \citep{r17} leaves the choice of the two cutoffs somewhat indeterminate and gives the force as:
\begin{equation}
{\bm F}=-16\pi^2(GM)^2m\log \left(\frac{\rmx}{r_{\rm min}} \right)\left[\int_0^V{\rm d}u\ u^2f(u)\right]\frac{\bm V}{V^3}, \label{chandra}
\end{equation}
where $G$ is the gravitational constant, $M$ is the perturber's mass, $m$ is the individual mass of a background medium star,  $V$ is the perturber's velocity, $f$  is the medium's velocity distribution function, $r_{\rm min}$ and $\rmx$  are the small and large distance cutoffs respectively. 
The small distance cutoff, $r_{\rm min}$,  is believed to be the larger of the perturber's radius and the distance related to the typical velocity in the ambient medium whereas the larger cutoff radius,  $\rmx$ ,  is usually taken as the size of the system (e.g. see \citep{r5}). A more accurate analysis of the gravitational scattering of a star in rectilinear motion in a collisionless infinite system \citep{r18}, shows that there is a natural large distance cutoff that equals the distance travelled by the perturber, $Vt$, where $t$ denotes the time the perturber takes crossing the medium. This cutoff arises for durations smaller than the crossing time because the far stars do not have enough time to completely deflect the perturber by the standard (asymptotic) angle that depends only on the distance to the perturber and its velocity but not on time. The farther the star, the more incomplete the deflection, leading to a scattering angle that depends on time. Equivalently, it can be shown that only stars with impact parameters smaller than $Vt$  substantially affect the perturber's velocity. This analysis was applied to accurately estimate the relaxation time in an infinite dilute star system \citep{r19}. When steady state is considered, the distance travelled by the perturber becomes comparable to the system's size and the upper cutoff, $\rmx$, is taken as the system's size. Similar prescriptions are applied to circular motion in spherically symmetric systems where it is shown that only the vicinity of resonances within the system contribute to dynamical friction \citep{r20} including the standard long and short distance cutoffs.

For a perturber moving in a gaseous medium, the small distance cutoff is taken as the perturber's accretion radius or physical size.  The large distance cutoff depends on the perturbation regime.  In steady state,  it is taken as the system's size \citep{r21,r22,r23} whereas in the time-dependent regime, in which the perturbation's duration is much smaller than the crossing time, it is given by the distance travelled by the sonic shockwave that the perturber excites in the medium \citep{r24,r25,r26,r27,r38} and \citep{r29} hereafter Paper I. Although the nature of the large distant cutoffs used to truncate the density perturbation is of different physical origin for collisionless and collisional media, the corresponding friction forces are similar \citep{r26}. 

In both collisionless and collisional media, dynamical friction was determined in the two limiting cases of steady state and the time-dependent regime. For the latter, the independence of dynamical friction of the medium's size is straightforward as the perturbed region that acts back on the perturber is much smaller than the ambient medium. The situation is much less clear for steady state in which the density enhancement raised by the perturber has reached the medium's boundary. Suppose for instance that a perturber is set in motion near the boundary of a gaseous medium such that the distance the sonic shockwave has to travel to reach the boundary is much smaller than the medium's size. Before the shockwave reaches the boundary, dynamical friction is in the time-dependent regime. However as the boundary is reached, the medium's full size cannot serve as a large distance cutoff. Instead, the proximity of the perturbed region to the medium's boundary will affect the way the density enhancement acts back on the perturber depending on how the boundary responds to an incoming shockwave. It is the object of this paper to find out how dynamical friction may be changed as the perturber's shockwave reaches the medium's boundary.  As such a task is quite difficult, we will make a number of simplifying assumptions aimed at illustrating that the classical behavior of dynamical friction may be changed drastically because of the presence of a boundary.  We assume that the perturber moves in a homogeneous gaseous medium along a rectilinear trajectory with uniform velocity near the medium's boundary that is taken to be planar and reflecting. The boundary's curvature is neglected as the medium's size is supposed to be large compared to the perturber's distance to the boundary. Wake reflection at a medium's boundary may occur at the edges of truncated disks perturbed by planetary or stellar companions \citep{r30,r31,r32,r42} as well as in numerical simulations of planet-disk interaction with no-outflow boundary conditions \citep{r33,r34,r35}. Interestingly, we shall show that most of the significant changes on dynamical friction come not from the reflected component but from the truncation of the shockwave incident on the boundary. To derive the friction force for this set-up, we shall use the general expression of dynamical friction derived in Paper I that allowed us to study the influence of a perturber's acceleration on dynamical friction.

The paper is organized as follows: in section 2, we recall the general expression of dynamical friction derived in Paper I and characterize how the medium's size affects dynamical friction in general.  In section 3, we derive the dynamical friction force for a perturber moving normal to the medium's boundary and show that perturbers are pushed away from the boundary and reach a finite terminal velocity.  In section 4, we examine dynamical friction for motion parallel to the boundary and show the appearance of an additional force component normal the trajectory  that pushes the perturber away from the boundary.  Section 5 includes a summary of main results and possible further refinements.  In Appendix A, we characterize the gas flow forced by the perturbers  in the configurations studied in  sections 3 and 4  by deriving the corresponding velocity field and density perturbation and linking them to the known results of the Bondi-Hoyle-Lyttleton accretion.

\section{General dependence of dynamical friction on the medium's size}
The reaction force exerted on a perturber, with a general potential $\phi_p$ and a general trajectory ${\bm\xi}(t)$, by its gravitational wake in a homogeneous gaseous medium, of unperturbed density $\rho_0$ and sound speed $c$, was derived in Paper I and reads:
\begin{eqnarray}
{\bm F} &=&\frac{{\cal H}(t)G\rho_0}{c^2}
\int_{\partial V[{\bm X}+{\bm \xi}(t)], |{\bm Y}|\leq ct}{\rm d}^3X\frac{{\rm d}^3Y}{|{\bm Y}|}  \  \ \rho_p({\bm X}-{\bm Y}+{\bm \Delta})\ {\bm \nabla} \phi_p({\bm X}),
\label{forc2A} \end{eqnarray}
where $\rho_p$ is the perturber's density distribution, $\rho_p=\nabla^2\phi_p/4\pi G$, ${\bm \Delta}={\bm \xi}(t)-{\bm \xi}(t-|{\bm Y}|/c)$ and ${\cal H}(t)$ is the Heaviside function. The spacial integration domain $\partial V({\bm x})$ is  defined by the medium's boundary as well as the lower distance cutoff in the perturber's vicinity such as its size or its corresponding accretion radius. The boundaries  are used to truncate the perturber's potential at small and large distances to avoid  the force's divergence. For a point-like perturber of mass $M$, the expression is simplified to:
\begin{equation}
{\bm F}=\frac{{\cal H}(t)(GM)^2\rho_0}{c^2}\int_{\partial V[{\bm y}+{\bm \xi}(t-r/c)], r\leq ct}r\sin\theta{\rm d}r\, {\rm d}\theta\, {\rm d}\varphi  \ \frac{{\bm y}-{\bm \Delta}}{|{\bm y}-{\bm \Delta}|^3},\label{force}
 \end{equation}
where  $r$, $\theta$ and $\varphi$ are the spherical coordinates of the vector ${\bm y}$ defined as the relative position within the medium with respect to the perturber's position at the retarded time $t-r/c$. It  may be written as ${\bm y}={\bm x}-{\bm \xi}(t-r/c)$ where ${\bm x}$ is the position of fluid element in the medium. This definition of ${\bm y}$ influences the force integral through the boundary conditions of the medium denoted  by $ \partial V$. The phase ${\bm \Delta}={\bm {\bm\xi}}(t)-{\bm {\bm\xi}}(t-r/c)$, the upper radial cutoff at the retarded sonic shockwave $r\leq ct$,  and the boundary conditions are all that is necessary to compute the friction force. In particular, as the force expression is independent of the density perturbation, no explicit expression for the density enhancement is  required for the force calculation (for more details see Paper I). The derivation of  expression (\ref{force}) assumes that the perturber is set in motion at $t=0$ a fact that introduces the size of the retarded sonic shockwave as an upper cutoff. This assumption is not restrictive as it 
allows us to study the time-dependent evolution of dynamical friction as well approximate steady state behaviour. It models a perturber that just entered the ambient medium  such as a star cluster or a dwarf galaxy that fall into a larger galaxy. It  also describes turning on a numerical simulation of dynamical friction \citep{r25,r26}. 

The formula (\ref{force}) allows us to see more clearly how the classical formula of steady state dynamical friction arises and how its dependence on the medium's size is erroneous.  Consider a perturber with uniform rectilinear motion that travels in a medium of a typical size $\rmx$. Writing the trajectory as ${\bm \xi}(t)=Vt {\bm e}_z$ where ${\bm e}_z$ is the unit vector along $z$, the phase ${\bm \Delta}={\cal M} r {\bm e}_z$ where ${\cal M}$ is the Mach number. The friction force is equally along  ${\bm e}_z$ and its expression is simplified as follows:
\begin{eqnarray}
{F}&=&  \frac{{\cal H}(t)(GM)^2\rho_0}{c^2}\int_{\partial V[{\bm y}+{\bm \xi}(t-r/c)], r\leq ct}r\sin\theta{\rm d}r\, {\rm d}\theta\, {\rm d}\varphi  \ \frac{r\cos \theta-{\Delta}}{|r^2+\Delta^2-2r\Delta\cos\theta|^{3/2}},\\
F&=&\frac{2\pi {\cal H}(t)(GM)^2\rho_0}{V^2}\int_{\partial V[{\bm y}+{\bm \xi}(t-r/c)], r\leq ct} \ \frac{{\rm d}r}
{r} \left[\frac{1-{\cal M}\tau}{({1+{\cal M}^2-2 {\cal M}\tau})^{1/2}}\right]_{\tau_{\rm min}}^{\tau_{\rm max}}, \label{force3}
\end{eqnarray}
where in the last expression the integrals over $\varphi$ and $\tau=\cos\theta$ were performed. The extremal values $ \tau_{\rm min}$ and $\tau_{\rm max}$ are determined from the boundary conditions. 

The condition on the medium's extent may be written as $|{\bm x}-{\bm \xi}(t)|\leq \rmx$ defining a sphere that comoves with the perturber.  Lastly, we supplement the integration domain with a lower cutoff $r_0$, usually taken to be the perturber's size or accretion radius, $2GM/V^2$, and implemented as   $|{\bm x}-{\bm\xi}(t)|>r_0$. Expressing these conditions in terms of the variables $r$ and $\tau$  that appear in equation (\ref{force3}) shows that we need to find the domains that satisfy: (i) $-1\leq\tau\leq1$, (ii) $r_0\leq r(1+{\cal M}^2-2 {\cal M}\tau)^{1/2}$, (iii) $r(1+{\cal M}^2-2 {\cal M}\tau)^{1/2}\leq \rmx$, and (iv) $r\leq ct$. The first three conditions yield the following integration domains:
\begin{eqnarray}
\frac{r_0}{|1+{\cal M}|}  <r<\frac{r_0}{|1-{\cal M}|} &\mbox{with}& \tau_{\rm min}=-1,\ \ \ \tau_{\rm max}=\frac{1+{\cal M}^2-(r_0/r)^2}{2{\cal M}}, \label{domain1}\\ 
\frac{r_0}{|1-{\cal M}|}  <r<\frac{\rmx}{|1+{\cal M}|}&\mbox{with}& \tau_{\rm min}=-1,\ \ \ \tau_{\rm max}=1,\label{domain2}\\ 
\frac{\rmx}{|1+{\cal M}|}  <r<\frac{\rmx}{|1-{\cal M}|} &\mbox{with}&   \tau_{\rm min}=\frac{1+{\cal M}^2-(\rmx/r)^2}{2{\cal M}} ,\ \ \ \tau_{\rm max}=1. \label{domain3}
\end{eqnarray}
The fourth condition depends on the distance $ct$ travelled by the retarded wavefront; it sets the dynamical friction regime:  time-dependent or steady state.  In the time dependent-regime, ($ct \ll \rmx/|1+{\cal M}|$), the sonic shockwave has not reached the medium's boundary whereas in the steady state regime, ($ct\gg \rmx/|1-{\cal M}|$), the peturbation has travelled the size of the medium. 
Calling $I$ the integral that appears in (\ref{force3}),  the integration for the time-dependent  regime extends over the two domains (\ref{domain1},\ref{domain2})  with $ct$ substituted for $\rmx/|1+{\cal M}|$ and $I$  is given as:
\begin{eqnarray}
I&=&-\log \left|\frac{1+{\cal M}}{1-{\cal M}}\right|+\frac{1}{2}\left[1+{\cal M}-|1-{\cal M}|\right]\left[1+{\rm sign}(1-{\cal M})\right] \nonumber\\&&
-\left[1-{\rm sign} (1-{\cal M})\right] \ \log\left|\frac{(1-{\cal M})ct}{r_0}\right|, \label{ostriker}
\end{eqnarray}
(see Paper I for a detailed calculation). This is the classical friction expression in the time-dependent regime \citep{r24,r25,r26,r27,r28}.  For the steady state regime, $|c-V|t$ is much larger than the size of the system. This condition ignores how the medium's boundary reacts to the incident shockwave and how the latter is truncated, reflected or transmitted by the boundary.  As $ct\geq\rmx/|1-{\cal M}|$, the condition from the propagation of the retarded wavefront (iv) becomes  irrelevant and  the three domains (\ref{domain1},\ref{domain2},\ref{domain3}) contribute finite terms to the force. The inner and middle domains yield the same result as in (\ref{ostriker}) except that $ct$ is replaced with $\rmx/|1+{\cal M}|$. The outer integration domain (\ref{domain3})  yields the term:
\begin{eqnarray}
I_{(\ref{domain3})}&=&   {\rm sgn}(1-{\cal M})\log \left|\frac{1+ {\cal M}}{1-{\cal M}}\right|-\frac{1}{2}\left[1+ {\cal M}-|1-{\cal M} |\right]\left[1+{\rm sgn}(1-{\cal M})\right]
\end{eqnarray}
which, when summed up with the terms of the first two domains, yields exactly: 
\begin{equation}
I =-\left[1-{\rm sgn} (1-{\cal M})\right] \ \log\left(\frac{\rmx}{r_0}\right). \label{steady1}
\end{equation}
By compensating for the contribution of the inner domain (\ref{domain1}), the force term  $I_{(\ref{domain3})}$ cancels dynamical friction for subsonic perturbers and yields a force on supersonic perturbers similar to the Chandrasekhar formula for collisionless systems. This is the classical result of  steady state dynamical friction \citep{r21,r22,r23}. The cancellation for subsonic perturbers happens only because of the choice of an outer boundary that is symmetric with respect to the perturber and comoves with it, i.e.  $|{\bm x}-{\bm \xi}(t)|\leq \rmx$; the two conditions are necessary. The change of behaviour from the time-dependent regime to steady state may be understood through the explicit expression of the perturbed density \citep{r24,r36}:
\begin{equation}
\varrho({\bm x},t)=\frac{GMk/c^2}{\left[(z-Vt)^2+R^2(1-{\cal M}^2)\right]^{1/2}} \ \mbox{where} \ \ 
k=\left\{\begin{array}{ll}
1& \mbox{if}\ R^2+z^2\leq (ct)^2,\\
2 & \mbox{if}\ {\cal M}>1,\ R^2+z^2> (ct)^2,  \\ &z+R\left({\cal M}^2-1\right)^{1/2}< Vt, \\ &\mbox{and}\ z{\cal M}< ct,  \\
0 &   \mbox{otherwise.}
\end{array}\right.
\label{density2}
\end{equation}
where $R=r\sin\theta$ is the cylindrical radius. Whereas the density perturbation is symmetric with respect to the perturber, the sonic wavefront of radius $ct$ that was launched at $t=0$ is not symmetric with respect to the perturber as it does not comove with it (Figure 1, top left panel). In the intermediate regime, when $ct\ll \rmx$ , the asymmetric part of the density perturbation inside the sonic wavefront results in dynamical friction.  As the wavefront reaches the comoving symmetric boundary, friction decreases and tends to zero when   $ct\gg \rmx$ as the density perturbation artificially acquires front-back symmetry  with respect to the perturber (Figure 1, bottom left panel).  However, this explanation that is based on the specifics of the density perturbation works only when uniform motion is considered as in general, motion is accelerated (or decelerated), as  by dynamical friction itself,  and the density enhancement generated by accelerated motion nowhere has front-back symmetry (Paper I). Discarding the unphysical assumption of truncating the force at large distances using  a comoving symmetric boundary (which is equivalent to implicitly assuming that perturber is at the center of a symmetric medium) will lead to significant changes in dynamical friction as suggested by  Figure 1 (right column) where the surfaces of equal density  are shown for a subsonic perturber in a medium with an arbitrary shaped boundary that does not comove with the perturber. As the sonic wavefront reaches the outer boundary, the density enhancement becomes asymmetric with respect to the perturber and friction is not only finite but will have a component perpendicular to the perturber's trajectory. In the following two sections, we examine a simpler boundary and show quantitatively that the properties of dynamical friction change significantly when the assumption of a comoving symmetric boundary is discarded. We also remark that most of our conclusions regarding  the use of a comoving symmetric boundary apply to collisionless systems. In effect, for rectilinear motion, the derivation of dynamical friction from the velocity impulse caused by a test star on the perturber is invariably centered around the latter. When steady state is considered, the cutoff $\rmx$ is used in the perturber-centered frame which is equivalent to assuming a comoving symmetric boundary. 

\section{Force for motion normal to the boundary}

As discussed in the previous section and shown in Figure 1, dynamical friction in the presence of a boundary unrelated to the perturber has parallel and normal components to the perturber's motion. To study each component separately, we choose perturbers that move normal to the boundary (this section) and parallel to the boundary (section 4).   We therefore consider an infinite medium with a reflecting planar boundary that occupies the space $z\geq 0$ and a perturber of trajectory ${\bm \xi}(t)=(Vt+d){\bm e}_z$ where ${\bm e}_z$ is the unit vector normal to the boundary. This setup models  dynamical friction on a perturber  whose outgoing wake has reached the nearest part of the medium's boundary or one that has just entered a gaseous medium. The assumption of a planar boundary is valid if the perturber is not near the physical center of the medium. No other forces  apply to the perturber in its motion in the medium. As the boundary is reflecting, the perturber's density enhancement will  have two components: the outgoing wake incident on the boundary and truncated by it, and a reflected wake. The reflection assumption serves two purposes, first as the large distance cutoff  (akin to the small distance cutoff) is commonly used to truncate the density enhancement and estimate dynamical friction, the outgoing wake truncated at the boundary will not be affected by reflection; the corresponding force would be the same for a non-reflecting boundary. Second, at little extra computational cost, we obtain the force caused by the reflected wake. Wake reflection at a medium's boundary occurs for instance at the edges of truncated disks perturbed by planetary or stellar companions \citep{r31,r32,r33} as well as in numerical simulations of planet-disk interaction with no-outflow boundary conditions \citep{r33,r34,r35}.

The sonic shockwave excited by the perturber  reaches the boundary at $t=d/c$. Prior to this time ($0\leq t\leq d/c$), the reaction force from the wake on the perturber is that of the time-dependent regime (\ref{ostriker}).  For later times, dynamical friction remains along the direction of the perturber's motion because the medium's boundary has a rotational symmetry with respect to the perturber's trajectory.   We consider separately the forces from the outgoing and reflected wakes and  assume that the perturber is not close to the boundary, i.e.  $d\gg (1+{\cal M})r_0/|1-{\cal M}|$.  Analytic expressions as well as portraits of the density perturbation and velocity fields are given in Appendix A. For the outgoing wake of a perturber moving away from the boundary, the geometrical truncation of the density perturbation will reduce dynamical friction. As the surfaces of equal density are symmetric with respect to the perturber, truncating part of the wake on the side opposite the perturber's motion (Appendix A, Figure 9) will effectively reduce the drag behind the perturber. In fact, we show below that the situation is more dramatic for slow subsonic perturbers as truncation reverses the direction of the force. The reflected wake remains behind the perturber for supersonic motion and overtakes the perturber at $t=2d/(1-{\cal M})$ for subsonic perturbers. Before this time and at all times for supersonic motion, the reflected wake is simply an additional density enhancement behind the perturber and increases the drag above that felt in an infinite medium because the same part of the wake is now closer to the perturber. After the reflected wake has overtaken a subsonic perturber, the amplitude of its drag decreases. We show below that its friction force may be reversed for slow subsonic motion in manner similar to that of the outgoing wake's force. Dynamical friction is derived for times $d/c \ll t \ll \rmx/c$. The first condition defines a local steady state with respect to the boundary nearest to the perturber and the second condition ensures that we do not follow the perturber's wakes (outgoing and reflected) as they reach and bounce off  the whole (closed) physical boundary of the medium. For a perturber moving towards the boundary,  dynamical friction is derived for times $t\leq d/V$ prior to the perturber's exit from the medium. Only subsonic motion is considered as the boundary ahead of a supersonic perturber does not affect dynamical friction.

\subsection{Force from outgoing wake}

We first assume that  the perturber moves away from the boundary (i.e. ${\cal M}\geq 0$). The boundary condition  $z\geq0$ written in terms of the variables of  expression (\ref{force3}) becomes $\tau\geq {\cal M}-\xi(t)/r$ with $-1\leq \tau\leq 1$.  The condition that the volume that contributes to the force be that inside the retarded sonic wavefront, i.e. $r\leq ct$, then implies that $\tau_{\rm min}=\xi(t)/(1+{\cal M})$ and $\tau_{\rm max}=1$ only for $t\geq d/c$. The integration domain related to the lower cutoff radius, $r_0$,   (\ref{domain1}) remains unchanged. For larger radii, we have:
\begin{eqnarray}
\frac{r_0}{|1-{\cal M}|}  \leq &r&\leq\frac{\xi(t)}{1+{\cal M}}\ \ \mbox{with}\ \  \tau_{\rm min}=-1,\ \ \ \tau_{\rm max}=1,\label{domain2i}\\ 
\frac{\xi(t)}{1+{\cal M}}  <&r&\leq c t\ \  \mbox{with}\ \    \tau_{\rm min}= {\cal M}- \frac{\xi(t)}{r},\ \ \ \tau_{\rm max}=1. \label{domain3i}
\end{eqnarray}
The domains (\ref{domain1}) and  (\ref{domain2i}) contribute force terms identical to those of equation (\ref{ostriker}) except that $ct$ is replaced with $\xi(t)/(1+{\cal M})$.  The third domain (\ref{domain3i}) related to the truncation of the sonic wavefront contributes the new term:
\begin{equation}
I_{(\ref{domain3i})}=\int_\frac{\xi(t)}{1+{\cal M}}^{ct}\left[\frac{{\rm sgn} (1-{\cal M})}{r} - \frac{(1-{\cal M}^2)r+{\cal M} \xi}{r^{3/2}\left[(1-{\cal M}^2) r + 2 {\cal M} \xi\right]^\frac{1}{2}}\right] {\rm d}r.
\end{equation}
Inspection of the boundary conditions associated with a perturber that moves towards the boundary (${\cal M}<0$) shows that the term $I_{(\ref{domain3i})}$ equally applies to $-1\leq {\cal M}\leq 0$ provided that the perturber does not leave the medium ($t< (d-r_0)/|V|$). Summing up the contributions of the three domains (\ref{domain1},\ref{domain2i},\ref{domain3i}) gives:
\begin{eqnarray}
I&=& \left(1+{\cal M}^2+2{\cal M}d/ct\right)^\frac{1}{2} +{\cal M}-1+\nonumber \\&& +   (1-{\cal M}^2)^\frac{1}{2} \log \left|\frac{ (d/ct+{\cal M})\left[1+\left(1-{\cal M}^2\right)^\frac{1}{2}\right]}{1+{\cal M}d/ct + \left[(1+{\cal M}^2 + 2{\cal M}d/ct)(1-{\cal M}^2)\right]^\frac{1}{2}}\right|\nonumber \\
&& -\log \left|\frac{d/ct+{\cal M}}{1-{\cal M}}\right| \ \ \ \mbox{for}\ \ \ |{\cal M}|<1, \label{incidentsub}\\
I&=& -3\left( 1+{\cal M}^2+2{\cal M}d/ct\right)^\frac{1}{2}    +3 {\cal M}+3+ \nonumber \\ &&+2({\cal M}^2-1)^\frac{1}{2}\left[\sin^{-1}\left(\frac{{\cal M}-1}{2{\cal M}}\right)^\frac{1}{2} -\sin^{-1}\left(\frac{{\cal M}^2-1}{2{\cal M}[{\cal M}+d/ct]}\right)^\frac{1}{2}\right] \nonumber \\
&&  -\log\left|\frac{({V}t+d)({\cal M}-1)ct}{r_0^2}\right|   \ \ \ \mbox{for}\ \ \ {\cal M}>1.\label{incidentsuper}
\end{eqnarray}
For the local steady state with respect to the boundary, $t\gg d/c$, the terms $d/ct$ disappear yielding dynamical friction as a function of ${\cal M}$ with the exception of the Coulomb logarithm for supersonic motion.
The force described by expressions  (\ref{incidentsub},\ref{incidentsuper}) is shown in Figure 2. As expected the truncation of the outgoing wake reduces the amplitude of the force for subsonic as well as supersonic perturbers that travel away from the boundary. For subsonic perturbers moving toward the boundary, the truncation of the wake region ahead of the perturber increases the dynamical drag as more weight is given to the density enhancement behind the perturber. For $-1<{\cal M}<0.43$, the truncation of the wake leads to the acceleration of the perturber away from the boundary.  Motion at ${\cal M}=0.43$ is in  stable equilibrium as for larger (smaller) velocities the force decelerates (accelerates) the perturber. Another interesting effect of the boundary's presence is the instability of perturbers at rest in the medium. While the integral $I$ (\ref{incidentsub}) vanishes as $|{\cal M}|$ tends to zero, the force which is proportional to $I/{\cal M}^2$ does not.
It is given as:
\begin{equation}
F(t\gg d/c,| {\cal M}|\ll 1)=-\frac{\pi{\cal H}(t) (GM)^2\rho_0\log |{\cal M}|}{c^2}. \label{instability}
\end{equation}
Perturbers at rest are unstable to small perturbations (Figure 5, bottom left panel)  and tend to be accelerated  toward the terminal velocity $V=0.43 c$. \footnote{It may be questioned whether the terminology `dynamical friction' applies to the force that accelerates perturbers instead of decelerating them. We choose to retain `dynamical friction' as the accelerating force derives from the same phenomenon and equation as standard dynamical friction. It is only the boundary's presence that is responsible for the behaviour change.} This instability originates in the fast  expansion of the truncated sonic wavefront with respect to the perturber's initial velocity ($c/V\gg 1$) giving most of the weight in the reaction force to the matter ahead of (behind) a perturber that moves away from (towards) the boundary.   

\subsection{Force from reflected wake}
In Appendix A, it is shown that the density perturbation associated with the wake reflected by a planar boundary is generated by a fictitious perturber whose trajectory is the mirror image of the actual perturber's motion with respect to the boundary, ${\bm x}\cdot {\bm n}=0$. The fictitious perturber's trajectory is   ${\bm \xi}^r(t)={\bm \xi}(t)-2[{\bm \xi}(t)\cdot{\bm n}] {\bm n}$, where ${\bm n}$ is the normal to the boundary.  The force, however, is applied at the perturber's position, ${\bm\xi}(t)$. Its expression is therefore given as:
\begin{eqnarray}
{\bm F}&=&\frac{{\cal H}(t)(GM)^2\rho_0}{c^2}\int_{\partial V({\bm x})}{\rm d}^3x\, {\rm d}u\ \frac{\delta
\left[u-t+|{\bm x}-{\bm \xi^r}(u)|/c\right]{\cal H}(u)}
{|{\bm x}-{{\bm\xi}^r}(u)|}\ \frac{{\bm x}-{{\bm\xi}}(t)}{|{\bm x}-{\bm {\bm\xi}}(t)|^3}, \nonumber\\
{\bm F}&=&\frac{{\cal H}(t)(GM)^2\rho_0}{c^2}\int_{\partial V[{\bm y}+{\bm \xi^r}(t-r/c)], r\leq ct} r\sin\theta{\rm d}r {\rm d}\theta{\rm d}\varphi \ \frac{{\bm y}-{\bm\Sigma}}{|{\bm y} -{\bm\Sigma}|^3}, \label{forcerref}
\end{eqnarray} 
where ${\bm y}={\bm x}-{\bm \xi^r}(t-r/c)$, $r$, $\theta$, $\varphi$ are its spherical coordinates  and ${\bm\Sigma}={\bm\xi}(t)-{\bm\xi}^r(t-r/c)$. For uniform rectilinear motion along ${\bm n}={\bm e}_z$, ${\bm \xi}^r(t)=-{\bm \xi}(t)=-(Vt+d){\bm e}_z$, $ {\bm y}={\bm x}+{\bm \xi}(t)$,  and  ${\bm \Sigma}=[2\xi(t)-{\cal M}r] {\bm e}_z$.  The force is along ${\bm e}_z$ and its amplitude is given as:
\begin{equation}
{F}= \frac{2\pi (GM)^2 {\cal H}(t)\rho_0}{V^2}\int_{\partial V} \ \frac{{\rm d}r}
{r} \left[\frac{{\cal M}^2(1-\Sigma \tau/r)}{(\Sigma/r)^2[1+(\Sigma/r)^2-2 \Sigma \tau/r]^{1/2}}\right]_{\tau_{\rm min}}^{\tau_{\rm  max}}.\label{forcer}\end{equation}
The three boundary conditions,  $|{\bm x}-{\bm\xi}(t)|\leq r_0$, $z\geq 0$ and $r\leq ct$ show that the force  is evaluated differently for subsonic and supersonic perturbers. There is only one integration domain as long as the wake does not overtake the perturber. This applies to supersonic perturbers at all times and only for $t\leq (2d- r_0)/(1-{\cal M})$ in the case of subsonic perturbers. This domain is given as:
\begin{eqnarray}
\frac{\xi(t)}{1+{\cal M}}  \leq &r& \leq ct\ \ \mbox{with}\ \  \tau_{\rm min}=\frac{\xi(t)}{r}-{\cal M},\ \ \ \tau_{\rm max}=1.\label{domainr1}
\end{eqnarray}
Calling $J$ the integral in equation (\ref{forcer}), these boundaries yield:
\begin{eqnarray}
J&=& \int_{\frac{\xi(t)}{1+{\cal M}}}^{ct} \ {{\rm d}r} \left[\frac{{\rm sgn}(1-\Sigma/r)}{r(\Sigma/r)^2}-\frac{(1-{\cal M}^2)r^2+ 3 \xi {\cal M} r - 2 \xi^2}{\Sigma^2r^{1/2}[(1-{\cal M}^2)r+ 2 \xi {\cal M}]^{1/2}}\right], \label{forcer2}\\
J&=& -\log \left|\frac{(1+2d/Vt)({\cal M}+1)}{(1+d/Vt)({\cal M}+2)}\right|-\frac{2(1-d/ct)}{({\cal M}+2)(1+2d/Vt)} +\frac{{\cal M}({\cal M}+1)}{{\cal M}+2}\nonumber \\ &&-\frac{{\cal M}(1+{\cal M}^2+2{\cal M}d/ct)^{1/2}}{{\cal M}+2d/ct}+ \nonumber\\
&&+\log\left|\frac{{\cal M}[1+({1+{\cal M}^2+2{\cal M}d/ct})^{1/2}]}{({\cal M}+2)[1-({1+{\cal M}^2+2{\cal M}d/ct})^{1/2}]}\right|+J_0, \label{j}\\
J_0&=& -({1-{\cal M}^2})^{1/2}\log\left|\frac{1+{\cal M}d/ct+[(1-{\cal M}^2)(1+{\cal M}^2+2{\cal M}d/ct)]^{1/2}}{({\cal M}+d/ct)\left[1+\left(1-{\cal M}^2\right)^{1/2}\right]}
\right|, \ \mbox{for}\ {\cal M}\leq 1,\label{j0sub}\nonumber\\
J_0&=& 2({\cal M}^2-1)^{1/2}\left[\sin^{-1}\left(\frac{{\cal M}^2-1}{2{\cal M}[{\cal M}+d/ct]}\right)^\frac{1}{2}-\sin^{-1}\left(\frac{{\cal M}-1}{2{\cal M}}\right)^\frac{1}{2}\right],  \  \mbox{for}\  {\cal M}\geq 1. \label{j0sup}\nonumber
\end{eqnarray}
In deriving these expressions we used the fact the first term in equation (\ref{forcer2}) is negative as $\Sigma>r$ for $r<2\xi(t)/(1+{\cal M})$. The wake completely overtakes the subsonic perturber after $t=(2d+ r_0)/(1-{\cal M})$. In this case the boundary conditions yield three domains as follows:
\begin{eqnarray}
\frac{\xi(t)}{1+{\cal M}}  \leq &r& \leq \frac{2\xi(t)-r_0}{1+{\cal M}}\ \ \mbox{with}\ \  \tau_{\rm min}=\frac{\xi(t)}{r}-{\cal M},\ \ \ \tau_{\rm max}=1,\label{domainr2}\\
\frac{2\xi(t)-r_0}{1+{\cal M}}  \leq &r& \leq \frac{2\xi(t)+r_0}{1+{\cal M}}\ \ \mbox{with}\ \  \tau_{\rm min}=\frac{\xi(t)}{r}-{\cal M},\ \ \ \tau_{\rm max}=\frac{r^2-r_0^2+\Sigma^2}{2 r \Sigma},\label{domainr3}\\
\frac{2\xi(t)+r_0}{1+{\cal M}}  \leq &r& \leq ct\ \ \ \  \ \ \  \ \ \ \ \ \  \mbox{with}\ \  \tau_{\rm min}=\frac{\xi(t)}{r}-{\cal M},\ \ \ \tau_{\rm max}=1.\label{domainr4}
\end{eqnarray}
The middle domain (\ref{domainr3}) evaluates the force near the perturber's location and it can be easily shown to  contribute a force term that vanishes as $r_0$ tends to zero. Calling $J_1$ the integral over the first and third domains (\ref{domainr2},\ref{domainr4}) defined by equation (\ref{forcer}), and noting that the sign of the first term in (\ref{forcer2}) is positive in the third domain, we find that the force integral on a subsonic perturber overtaken by the reflected wake is:
\begin{equation}
J_1=J+{2}\log\left|\frac{({\cal M}+1)(1+2d/Vt)}{2(1+d/Vt)}\right|+\frac{2{\cal M}(1-{\cal M}-2d/ct)}{{\cal M}+2d/ct}.\label{j1}
\end{equation}
As for the outgoing wake, inspection of the boundary conditions for a perturber moving towards the boundary shows that the expressions (\ref{j},\ref{j0sub},\ref{j1}) are valid for $|{\cal M}|<1$. The force from the reflected wake is shown in Figure 3. For supersonic and subsonic perturbers ahead of the wake, the reflected wake increases the dynamical drag. For larger Mach numbers, the contribution of the reflected wake is much smaller than  that of the outgoing wake. The force amplitude is reduced after the reflected wake overtakes the subsonic perturber. Below ${\cal M}=0.27$, the reflected wake accelerates subsonic perturbers in the same manner as the outgoing wake. Perturbers at rest are equally unstable and the local steady state force from the reflected wake for slow subsonic perturbers is equal to that from the outgoing wake (\ref{instability}).

The reaction force from both outgoing and reflected wakes is shown in Figure 4. Dynamical fiction is reversed at ${\cal M}=0.37$, a velocity that defines the effective stable rest state. The action of the reflected wake makes the dynamical drag  beyond ${\cal M}=0.57$ stronger than that in an infinite medium  peaking near ${\cal M}=0.8$ at $20\%$ above that of equation (\ref{ostriker}). Zero-velocity perturbers are unstable and are subjected to a force twice that from the outgoing wake (\ref{instability}).  As a result of the boundary's presence, supersonic perturbers feel a smaller drag force that is about $10\%$ below the value of the force in an infinite medium. The peak of dynamical drag remains near Mach 1. 

\section{Force for motion parallel to the boundary}
For a perturber moving in an infinite homogeneous medium occupying the space $x\geq 0$, parallel to its reflecting boundary, the gravitational wake will be truncated only on one side parallel to the perturber's trajectory ${\bm\xi}(t)=d {\bm e}_x+Vt {\bm e}_z$ (Appendix A, Figure 10). Motion parallel to the boundary is therefore subject to an additional component normal to the trajectory unlike motion normal to the boundary which results only in a force along the perturber's trajectory. As $t$ becomes much larger than $d/c$, the forces from the outgoing and reflected wakes along the perturber's trajectory will tend to  similar values close to half that of expression  (\ref{ostriker}) as the wake is approximately  cut by half (Figure 10). The total force along the trajectory will therefore tend to its value in an infinite medium. The force normal to the trajectory however will increase with the size of the perturbed region and force the perturber to move away from the boundary. 

\subsection{Force from outgoing wake}
The force from the outgoing wake is derived using the general expression (\ref{force}). With our choice of medium with $x\geq 0$ and ${\bm\xi}(t)=d {\bm e}_x+Vt {\bm e}_z$,  ${\bm \Delta}={\cal M}r {\bm e}_z$. The main difference in the derivation of the force with respect to sections 3 and 5 is the integration domain of the variable $\varphi$ that not longer runs from 0 to $2\pi$ as rotational symmetry is broken for $r>d$. Dividing the integration domain over $r$ into $r_0/|1+{\cal M}|\leq r\leq d$ and $d\leq r\leq ct$ and writing ${\bm F}=F_x{\bm e}_x+F_z{\bm e}_z$, the force related to the former domain $F_x(r_0/|1+{\cal M}|\leq r\leq d)=0$ whereas $F_z(r_0/|1+{\cal M}|\leq r\leq d)$ is given by equation (\ref{ostriker}) where $ct$ is replaced with $d$. In writing this we assumed that $d\gg (1+{\cal M})r_0/|1-{\cal M}|$ so that the perturber is not too close to the boundary. For the part related to the domain $d\leq r\leq ct$, the force expressions may be written  as: 
\begin{eqnarray}
F_x(d\leq r\leq ct)&=&
\frac{{\cal H}(t)(GM)^2\rho_0}{c^2}\int_{\partial V} \frac{\sin^2\theta {\rm d}\theta{\rm d}r\ [\sin\varphi]^{\varphi_{\rm max}}_{\varphi_{\rm min}}}{r\left(1+{\cal M}^2-2 {\cal M} \cos\theta\right)^{3/2}}, \label{forcepx}\\
F_z(d\leq r\leq ct)&=&\frac{{\cal H}(t)(GM)^2\rho_0}{c^2}\int_{\partial V} \frac{(\cos\theta-{\cal M})\sin\theta {\rm d}\theta{\rm d}r}{r\left(1+{\cal M}^2-2 {\cal M} \cos\theta\right)^{3/2}}\ \left[{\varphi_{\rm max}}-{\varphi_{\rm min}}\right],\label{forcepz}
\end{eqnarray}
where the integration over $\varphi$ was performed.  In the variables of equation (\ref{force}),  the boundary condition $x\geq 0$ becomes $r\sin\theta\cos\varphi\geq -d$. Denoting, $\cos\varphi_d=-d/r\sin\theta$ and $\sin\theta_d=d/r$, the boundary condition yields the angles $\varphi_{\rm min}$ and $\varphi_{\rm max}$ as functions of $\theta$ and $r$ as follows:
\begin{eqnarray}
 \varphi_{\rm min}=0, \ \varphi_{\rm max}=2\pi &\mbox{for}& 0  \leq\theta\leq\theta_d\ \ \ \mbox{and}\ \ \ 
 \pi-\theta_d  \leq\theta\leq\pi, \nonumber\\
\varphi_{\rm min}=-\varphi_d, \ \ \ \varphi_{\rm max}=\varphi_d&\mbox{for}&  \theta_d  \leq\theta\leq\pi-\theta_d.\nonumber
\end{eqnarray}
Substituting these expressions into equations (\ref{forcepx},\ref{forcepz}) yields:
\begin{eqnarray}
F_x(d\leq r\leq ct)
&=&\frac{{\cal H}(t)(GM)^2\rho_0}{c^2}\int_{r=d}^{ct}\int_{\theta=\theta_d}^{\pi-\theta_d} \frac{2\sin\theta {\rm d}\theta{\rm d}r\left(r^2\sin^2\theta-d^2\right)^{1/2}}{r^2\left(1+{\cal M}^2-2 {\cal M} \cos\theta\right)^{3/2}}\ \label{forcepx2} \\
&&+\frac{{\cal H}(t)(GM)^2\rho_0}{c^2}\int_{r=d}^{ct}\int_{\theta=0}^{\theta_d} \frac{2\pi\sin^2\theta {\rm d}\theta{\rm d}r}{r} \left[\left(1+{\cal M}^2-2 {\cal M} \cos\theta\right)^{-3/2}\right. \nonumber \\ && \left.+\left(1+{\cal M}^2+2 {\cal M} \cos\theta\right)^{-3/2}\right],\nonumber\\
F_z(d\leq r\leq ct)&=&\frac{{\cal H}(t)(GM)^2\rho_0}{c^2} \int_{r=d}^{ct}\int_{\theta=\theta_d}^{\pi-\theta_d} \frac{2(\cos\theta-{\cal M})\sin\theta {\rm d}\theta{\rm d}r}{r\left(1+{\cal M}^2-2 {\cal M} \cos\theta\right)^{3/2}}\ \cos^{-1}\left(\frac{-d}{r\sin\theta}\right)           \nonumber \\
&&+        \frac{{\cal H}(t)(GM)^2\rho_0}{c^2}  \int_{r=d}^{ct}\int_{\theta=0}^{\theta_d} \frac{2\pi\sin\theta {\rm d}\theta{\rm d}r}{r}\     \left[\frac{\cos\theta-{\cal M}}{\left(1+{\cal M}^2-2 {\cal M} \cos\theta\right)^{3/2}}\right. \nonumber \\ && \left.-\frac{\cos\theta+{\cal M}}{\left(1+{\cal M}^2+2 {\cal M} \cos\theta\right)^{3/2}}\right] .\label{forcepz2}                                                                                                                                                                                                                                                                                                                                                                                                             
\end{eqnarray}
These expressions cannot be reduced analytically if local steady state is not assumed. To follow the force's time evolution as it reaches steady state, the expressions are integrated numerically, combined with the force term associated with the domain $r_0/|1+{\cal M}|\leq r\leq d$ and shown in Figure 5. As expected, it is apparent on the magnitude of $F_z$ for subsonic perturbers and on the slope of $F_z(t)$ for supersonic perturbers that the force along the trajectory $F_z$  tends to half the value given by the force on a perturber in a infinite medium (\ref{ostriker}). We note that while this result is straightforward when looking at Figure 10, it could not have been guessed from the formula used for an infinite medium (\ref{ostriker}) as only the constant unperturbed density represents  the medium in that expression and it is not reduced by half if there is a boundary. 
The force normal  to the direction of motion, $F_x$, appears as soon as the boundary breaks the rotational  symmetry of the wake incident on it after $t=d/c$. The force from the reflected wake always pushes the perturber away from the boundary and increases linearly with the size of the perturbed region which is proportional to $\log (ct)$. It is strongly peaked at  ${\cal M}=1$ where it reaches its finite maximum.  In the local steady state regime, $ct\gg d$, the normal force is given by the first term of equation (\ref{forcepx2}) which may be simplified to:
\begin{equation}
F_x=\frac{2{\cal H}(t)(GM)^2\rho_0 \log (ct/d)}{V^2(1+{\cal M})} \left\{(1+{\cal M}^2)\  {\rm K}\left[\frac{4{\cal M}}{(1+{\cal M})^2}\right]-(1+{\cal M})^2 \ {\rm E}\left[\frac{4{\cal M}}{(1+{\cal M})^2}\right] \right\},\label{fxlimss}
\end{equation}
where K and E are the complete elliptic integrals of the first and second kind respectively. This expression is not valid at ${\cal M}=1$ as it gives an infinite force (the maximum amplitude of $F_x$ (\ref{forcepx2}) at ${\cal M}=1$ grows faster than $(\log ct/d)^2$ and its value becomes infinite in an expansion  in terms of $d/ct$). For slow subsonic perturbers, ${\cal M}\ll 1 $, the force is independent of velocity and may be written as:
\begin{equation}
F_x=\frac{\pi{\cal H}(t)(GM)^2\rho_0 \log (ct/d)}{c^2}\label{fxlimss1}.
\end{equation}
For fast  supersonic perturbers, ${\cal M}\gg 1$, the normal force is given as:
\begin{equation}
F_x=\frac{\pi{\cal H}(t)(GM)^2c\rho_0 \log (ct/d)}{V^3}\label{fxlimss2}.
\end{equation}
This expression of  $F_x$ bears some resemblance to that of $F_z$ (\ref{ostriker})  except that (i) the former's amplitude is smaller than the latter's by a factor ${\cal M}$, (ii) it is the distance to the boundary that takes on the role of $r_0$ in the Coulomb logarithm, (iii) the size of the perturbed region is $ct$ instead of $Vt$ for supersonic perturbers.

\subsection{Force  from reflected wake}
The force from the reflected wake is derived using the general expression (\ref{forcerref}) with  ${\bm \Sigma}={\bm \xi}(t)-{\bm \xi}^r (t-r/c)=2 d {\bm e}_x+{\cal M}r {\bm e}_z$.  In the variables of equation (\ref{forcerref}), the condition $x\geq 0$ becomes $r\sin\theta\cos\varphi\geq d$  implying that $d\leq r$, $\theta_d\leq \theta\leq \pi-\theta_d$ and $\varphi_d-\pi\leq \varphi\leq \pi-\varphi_d$.  
The boundary condition related to the lower cutoff distance $|{\bf y}-{\bf \Sigma}|\geq r_0$ shows that there is a range in radius between $r_\pm=(2d\sin\theta\pm d_1)/(1+{\cal M}^2-2{\cal M}\cos\theta)$ with $d_1^2=r_0^2(1+{\cal M}^2-2 {\cal M} \cos\theta)-4 d^2(\cos\theta-{\cal M})^2$ where
\begin{equation}
\cos\varphi\leq \frac{r^2(1+{\cal M}^2-2 {\cal M} \cos\theta)+4d^2-r_0^2}{4rd\sin\theta}.
\end{equation}
This condition is the counterpart of the middle domain (\ref{domainr3}) associated with perturbers moving normal to the boundary that samples the force from the reflected wake in the perturber's vicinity. In the limit where $r_0$ tends to zero it can be shown easily that this condition does not affect the boundary values of the variable $\varphi$.
The force from the reflected wake is therefore given as:
\begin{eqnarray}
F_x
&=&\!\!\!\!\frac{{\cal H}(t)(GM)^2\rho_0}{c^2}\!\!\!\int_{r=d}^{ct}\!\int_{\theta_d}^{\pi-\theta_d} \!\!\!\!\!\int_{\varphi=0}^{\varphi=\pi-\varphi_d}\!\!\!\!\!\!\frac{2r(r\sin\theta\cos\varphi-2d) \sin\theta {\rm d}\theta{\rm d}r{\rm d}\varphi}
{[r^2(1+{\cal M}^2-2 {\cal M} \cos\theta)+4d^2-4rd\cos\varphi\sin\theta]^{3/2}}\nonumber \\
&&\\
F_z&=& \!\!\!\!    \frac{{\cal H}(t)(GM)^2\rho_0}{c^2}\!\!\!\int_{r=d}^{ct}\!\int_{\theta_d}^{\pi-\theta_d} \!\!\!\!\!\int_{\varphi=0}^{\varphi=\pi-\varphi_d}\!\!\!\!\!\!\frac{2r^2(\cos\theta-{\cal M})\sin\theta {\rm d}\theta{\rm d}r{\rm d}\varphi}
{[r^2(1+{\cal M}^2-2 {\cal M} \cos\theta)+4d^2-4rd\cos\varphi\sin\theta]^{3/2}} \nonumber \\
&&                                                                                                                                                                                                                                                                                                                                                                                                       
\end{eqnarray}
Like those of the previous section, these expressions cannot be reduced further analytically. They are integrated numerically to see how the force reaches local steady state and shown in Figure 6. The component along the trajectory, $F_z$, tends to a value slightly smaller than half the value of (\ref{ostriker}) because of the position shift $d$ with respect to the outgoing wake -- for supersonic motion, the factor half is seen on the slope of the force.  Local steady state is reached after a transitory period where the reflected wake is located in the vicinity of the boundary and does not cross the perturber's trajectory. This is the reason why $F_x$ initially pulls the perturber to the boundary before reaching its asymptotic positive value that is proportional to $\log ct$. Like the force from the outgoing wake, $F_x$ is maximal at ${\cal M}=1$.    The analytical expressions of  the normal force from the reflected wake in the local steady state regime, $ct\gg d$,  are identical to those of the outgoing wake (\ref{fxlimss},\ref{fxlimss1},\ref{fxlimss2}).

The reaction force from both outgoing and reflected wakes is shown in Figure 7. The force along the trajectory is comparable in amplitude to that in an infinite medium except in the neighborhood  of Mach 1 where its is larger by about 15\%, a trend that is opposite to that of the friction force acting  on a perturber moving normal to the boundary. The force normal to the trajectory increases linearly with $\log (ct/d)$  (for ${\cal M}\neq 1$) and  is maximal at Mach 1.  For ${\cal M}\leq 1$, $F_x > |F_z|$ whereas for  ${\cal M}\geq 1$, $F_x \ll |F_z|$. We note that like for motion normal to the boundary, the gravitational wakes tend to push the perturber away from the boundary. 

\section{Conclusion}
We set out to examine the effect of the presence of a boundary on dynamical friction. We showed that the expressions of the force in  infinite (\ref{ostriker}) and finite  (\ref{steady1})  media may not be used in the presence of a boundary such as the medium's outer and inner limits or any other internal edge.  In particular, dynamical friction along the direction of motion may be reversed and a new drag component perpendicular to the boundary  is present and forces the perturber to move away from the medium's edge.  These results are of particular relevance for the numerical simulation of dynamical friction once the perturber's wake reaches the boundaries of the physical system or the simulation box as for instance  in problems related to disk-planet interactions and  galaxy merging where timescales are observed to differ significantly from the prediction of the Chandrasekhar formula \citep{r43}.

We have considered separately motion normal and parallel to the boundary to isolate the effects of the instability of zero-velocity perturbers, the reversal of dynamical friction, and  normal dynamical friction. Whereas qualitatively accurate for a perturber on a trajectory inclined by an arbitrary angle with respect to the boundary, these effects may not be added quantitatively linearly because dynamical friction depends on the phase vectors ${\bm \Delta}$ and ${\bm \Sigma}$. This is why it is not straightforward to carry over the first two effects to rotational motion in bounded systems (such as disk-planet interactions) as they would require the perturber's orbit to have some eccentricity and hence a velocity that makes a time-dependent angle with the boundary. Normal dynamical friction however readily applies to circular motion in a spherical system. The dependence of the normal force on the distance to the boundary is translated for a  perturber in circular motion around an attracting center as a radial force that depends on the orbital radius and the location of  the medium's edge for subsonic perturbers, and additionally on the velocity for supersonic perturbers. This time-dependent force affects the perturber's energy but not its angular momentum and is most efficient where motion is resonant with the sound speed.  

In our derivation of dynamical friction, we neglected the boundary's curvature. It is straightforward that for a perturber on a normal trajectory that moves away from a convex boundary, the friction force will be further reduced with respect to  that given by (\ref{incidentsub},\ref{incidentsuper}) as the wake behind the perturber is truncated further (as in Figure 1). 

The current analysis may also serve as a guide to examine the effect of the ambient medium's boundary in collisionless systems. As mentioned earlier, the derivation of dynamical friction from the velocity impulse caused by a test star on the perturber is centered physically around the latter \citep{r5}. When steady state is considered, the large distance cutoff is applied in the perturber-centered frame which is equivalent to assuming a comoving symmetric boundary.  Although they differ in magnitude, the  friction forces in a gaseous medium and a collisionless star system share a strong similarity in the way they depend on the Mach number \citep{r26}. Such similarity suggests that the conclusions we found  here will  hold for collisionless systems.

 \appendix
\section{Gas flow near the medium's boundary}
In this Appendix, we characterize the flow in the presence of a reflecting planar boundary through  the density enhancement and the velocity field. Prior to discussing the case of a medium with a boundary, we first determine the density enhancement and velocity field in an infinite medium. Much work has been devoted to the study of a similar configuration that of the Bondi-Hoyle-Lyttleton accretion problem (for a review see \cite{r37}). However most analytical studies of the flow have always dealt only with the ballistic flow for supersonic perturbers therefore excluding the effect of pressure in the medium. As explained below the ballistic flow occurs outside the wake region and is therefore irrelevant to dynamical friction because it is not accompanied by a density enhancement. In the context of Bondi-Hoyle-Lyttleton accretion with pressure, the analytical studies of the flow have characterized the velocity only in the vicinity of the singular and asymptotic points \citep{r38,r39,r40}.

The medium's response to the excitation of a point-like perturber of gravitational potential $\phi_p=-GM/r$   that is set in motion  at time $t=0$ is described by the standard fluid equations:
\begin{equation}
\partial_t\rho+{\bm \nabla} \cdot (\rho {\bm v})=0,\ \ \ \partial_t {\bm v}+({\bm v}\cdot{\bm \nabla}) {\bm v}=-\frac{1}{\rho}{\bm \nabla} p-{\cal H}(t){\bm \nabla}{\phi_p},
\end{equation}
where ${\bm v}({\bm x},t)$ and  $\rho({\bm x},t)$ are the perturbed velocity and density fields, $p$ is the pressure, and ${\cal H}(t)$ is the Heaviside function. Prior to the perturber's motion, the gaseous medium is assumed to be at rest $[i.e. {\bm v}({\bm x},t)=0]$, homogeneous of density $\rho_0$, pressure $p_0$ and isothermal with a sound speed $c=(p_0/\rho_0)^{1/2}$. The linearization of medium's equation in the vicinity of the rest state yields: 
\begin{equation}
\partial_t\varrho+c{\bm \nabla} \cdot { \bm{\nu}}=0,\ \ \ 
\frac{1}{c}\partial_t {\bm \nu}+{\bm \nabla} \varrho= -\frac{{\cal H}(t)}{c^2}{\bm \nabla} \phi_p,  \label{momentumeq}
\end{equation}
where we used $\rho=\rho_0(1+\varrho)$ and ${\bm v}=c {\bm \nu}$. 
The combined equations in turn yield the forced sound wave equation satisfied by the density perturbation $\varrho$ as: 
\begin{equation}
\nabla^2\varrho-\frac{1}{c^2}\,
\partial_t^2\varrho=-\frac{{\cal H}(t)}{c^2}\nabla^2\phi_p \label{wavA}.
\end{equation}
The solution of this equation is known in electromagnetism theory as the Li\'enard-Wiechert potential and represents the electric potential of a moving electric charge distribution.  It is written in integral form using the retarded Green function as:
\begin{eqnarray}
\varrho({\bm x},t)&=&\frac{1}{4\pi c^2}\int_{-\infty}^{+\infty}{\rm d}^3x^\prime {\rm d}u\ \frac{\delta
\left[u-t+|{\bm x}-{\bm x}^\prime|/c\right] \ \nabla^2\phi_p[{\bm x}^\prime - {\bm \xi}(u)]\ {\cal H}(u)}
{|{\bm x}-{\bm x}^\prime|}, \nonumber\\
\varrho({\bm x},t)&=&\frac{GM}{c^2}\int_{-\infty}^{+\infty} \frac{\delta
\left[u-t+|{\bm x}-{\bm \xi}(u)|/c\right] {\cal H}(u)}
{|{\bm x}-{\bm \xi}(u)|}\ {\rm d}u.  \label{densA}
\end{eqnarray} 
The momentum equation (\ref{momentumeq}) shows that the velocity, ${\bm \nu}$, can be written in terms of a potential function  $\psi$ as ${\bm \nu}={\bm \nabla} \psi$ given that the medium is at rest prior to the perturber's motion. The potential $\psi$ satisfies the equation:
\begin{equation}
\partial_t \psi({\bm x},t)= -\frac{1}{c}[c^2\varrho({\bm x},t)+ \phi_p({\bm x},t)],  
\end{equation}
which upon integration yields:
\begin{eqnarray}
\psi({\bm x},t)&=& -\frac{1}{c}\int_0^t [c^2\varrho({\bm x},w)+ \phi_p({\bm x},w)]\,{\rm d} w =\frac{GM}{c} (\psi_\varrho+\psi_{\rm ball}),\\
\psi_\varrho({\bm x},t)&=&  -\int_{u=-\infty}^{\infty} \int_{w=0}^t{\rm d}u\, {\rm d}w \frac{\delta
\left[u-w+\left(R^2+\zeta(u)^2\right)^{1/2}/c\right] } {\left[R^2+\zeta(u)^2\right]^{1/2}}\ {\cal H}(u)\nonumber \\ &=&-\int_{0\leq u+\left(R^2+\zeta^2\right)^{1/2}/c \leq t, \ u\geq0 }\frac{{\rm d}u} {\left[R^2+\zeta(u)^2\right]^{1/2}},\nonumber\\
\psi_\varrho({\bm x},t)&=&\left[\frac{1}{V}\log\left(z-Vu+\left[R^2+(z-Vu)^2\right]^\frac{1}{2}\right)\right]_{0\leq u+\left(R^2+\zeta^2\right)^{1/2} /c\leq t, \ u\geq0 },\\
\psi_{\rm ball}({\bm x},t)&=& \int_{0}^{t}\frac{ {\rm d}w } {\left[R^2+\zeta(w)^2\right]^{1/2}}=-\frac{1}{V}\log\left(\frac{z-Vt+\left[R^2+(z-Vt)^2\right]^{1/2}}{z+\left[R^2+z^2\right]^{1/2}}\right), \label{psiout}
\end{eqnarray}
where $\zeta(u)=z-\xi(u)$ and is $\psi_{\rm ball}$ the ballistic potential associated with the pressureless flow \citep{r41}. To determine the boundaries of the  density perturbation's contribution $\psi_\varrho$   to the velocity potential $\psi$, we find the roots of $cu+\sqrt{R^2+\zeta^2}\leq ct$ subject to the conditions $0\leq u\leq t$. These roots are given as:
\begin{equation}
u_\pm=\frac{ct-{\cal M}z\pm \left[(ct-{\cal M}z)^2-(1-{\cal M}^2)(ct^2-r^2)\right]^{1/2}}{c(1-{\cal M}^2)}.
\end{equation}
For subsonic motion and  outside the sonic wavefront, $r=ct$, there are no roots  as there is no density perturbation. This implies  that $\psi_\varrho=0$ and  $\psi=\psi_{\rm ball}$. Inside the sonic wavefront, $r<ct$, $u_{-}$ is the only possible root so that both  $\psi_\varrho$ and $\psi_{\rm ball}$ contribute to $\psi$ and yield:
\begin{equation}
\psi({\bm x},t)=-\frac{1}{V}\log\left(\frac{z-Vt+\left[R^2+(z-Vt)^2\right]^{1/2}}{z-Vu_{-}+\left[R^2+(z-Vu_{-})^2\right]^{1/2}}\right)\equiv \psi_{\rm sonic}({\bm x},t).\label{psisub}
\end{equation}
For supersonic motion, there are three domains: (i) outside the sonic wavefront and the Mach cone: $ct\leq r$, $ct\leq (z+R\sqrt{{\cal M}^2-1})/{\cal M}$ and $ct/{\cal M}\leq z\leq {\cal M} ct$, (ii) inside the sonic wavefront: $r<ct$,   (iii) outside the sonic wavefront but inside the Mach cone:  $(z+R\sqrt{{\cal M}^2-1})/{\cal M}\leq ct\leq r$ and $ct/{\cal M}\leq z\leq {\cal M} ct$. Outside the sonic wavefront and the Mach cone,  no roots are possible, $\psi_\varrho=0$  and $\psi=\psi_{\rm ball}$. Inside the sonic wavefront, only $u_{-}$ is possible  and $\psi=\psi_{\rm sonic}$.   
Inside the Mach cone where $(z+R\sqrt{{\cal M}^2-1})/{\cal M}\leq ct\leq r$ and $ct/{\cal M}\leq z\leq {\cal M} ct$, the two roots $u_{\pm}$ are possible and the velocity potential is given as:
\begin{eqnarray}
\psi({\bm x},t)&=&-\frac{1}{V}\log\left[\frac{\left(z-Vu_{+}+\left[R^2+(z-Vu_{+})^2\right]^{1/2}\right)\left(z-Vt+\left[R^2+(z-Vt)^2\right]^{1/2}\right)}{\left(z-Vu_{-}+\left[R^2+(z-Vu_{-})^2\right]^{1/2}\right)\left(z+\left[R^2+z^2\right]^{1/2}\right)}\right]\nonumber \\&&\equiv \psi_{\rm Mach}({\bm x},t). \label{psisuper}
\end{eqnarray}
The calculated flow (\ref{psiout},\ref{psisub},\ref{psisuper}) is shown in Figure 8. The ballistic flow outside the wake region, $\psi_{\rm ball}$, is finite despite the absence of a density enhancement. This results from the infinite interaction range of the gravitational potential $\phi$. This part of the flow however does not give rise to dynamical friction. The wavefront as a whole propagates as a shock in the gaseous medium. Inside the sonic wavefront, the velocity  vanishes along the perturber's trajectory. For supersonic motion, whereas the velocity potential is continuous at the sonic wavefront $r=ct$,  the velocity is discontinuous  reflecting the discontinuity of the density $\varrho$ (\ref{density2}). The charateristics of the flow given by $\psi_{\rm sonic}$, and  $\psi_{\rm Mach}$ agree with those obtained from the numerical simulation of the nonlinear fluid equations (\citep{r25}, Figure 1). The velocity potentials  we derived does not agree with the expressions of the velocity field quoted without derivation by \citep{r36}. 

We now turn to the problem of a perturber on a trajectory ${\bm \xi}(t)$ in a medium with a infinite planar boundary defined as ${\bm x}\cdot {\bm n}=0$ where ${\bm n}$ is its  constant normal vector. The coordinate frame is chosen so that the boundary contains the origin. The medium fills  the space where ${\bm x}\cdot {\bm n}\geq 0$.  For the outgoing wake, the density and velocity perturbations are truncated geometrically by the boundary as shown in Figure 1 (bottom right panel). To obtain the density and velocity perturbations of the reflected wake, we use the property that normal fluid motion vanishes on the boundary. This may be written as  ${\bm\nu}({\bm x},t)\cdot {\bm n}=0$  for  ${\bm x}\cdot {\bm n}=0$. The equations of the density perturbation and the velocity potential  to be solved for ${\bm x}\cdot {\bm n}\geq 0$ are given by:
 \begin{eqnarray}
&& \nabla^2\varrho_1-\frac{1}{c^2}\, \partial_t^2\varrho_1=-\frac{1}{c^2}\nabla^2\phi_p,\ \ \ 
\psi_1= -\frac{GM}{c}\int_0^t (c\varrho_1+ \phi_p) {\rm d} \tau, \nonumber \\ &&\mbox{and} \ \ {\bm\nabla}\psi_1 \cdot {\bm n}=0 \ \ \ \mbox{at} \ \ {\bm x}\cdot {\bm n}=0.\label{dalembert}
\end{eqnarray}
This problem can be solved by the method of images used  in electromagnetism theory. The system that consists of a perturber and a boundary is equivalent to that of two identical perturbers placed on each side of the boundary and whose trajectories are mirror images of one another with respect to the boundary. To see this we note first that we seek the solution of the forced wave equation (\ref{dalembert}) in the region ${\bm x}\cdot {\bm n}\geq0$.  The second perturber,  whose trajectory ${\bm \xi}^r(t)={\bm \xi}(t)-2[{\bm \xi}(t)\cdot{\bm n}] {\bm n}$, is located in the region ${\bm x}\cdot {\bm n}<0$.  Consequently, its corresponding density perturbation satisfies the force free wave equation in the region ${\bm x}\cdot {\bm n}\geq0$ and  could  simply be added to the wave equation. Second, the trajectories of the two perturbers, ${\bm\xi}(t)$ and  ${\bm \xi}^r(t)$,  being mirror images of one another with respect to the boundary, the normal components of the respective velocity fields  are antisymmetric and sum up to zero at the boundary. For a perturber moving inside the medium normal to its boundary initially at a distance $d$ from it, we choose ${\bm n}={\bm e}_z$. We then have ${\bm \xi}(t)=(Vt+d){\bm e}_z$,  ${\bm \xi}^r(t)=-(Vt+d){\bm e}_z$ and the density perturbation and velocity potential are given as:
\begin{eqnarray}
\varrho_1&=& {\cal H}(z)\left[\varrho(R,z-d,V,t)+\varrho(R,z+d,-V,t)\right],\\
\psi_1&=&{\cal H}(z)\left[\psi(R,z-d,V,t)+\psi(R,z+d,-V,t)\right],
\end{eqnarray}
where $\varrho$ and $\psi$ are given respectively by equations (\ref{density2}) and (\ref{psiout}, \ref{psisub}, \ref{psisuper}). The first term in each equation corresponds to the geometrically truncated outgoing wake. The second term  corresponds to the reflected wake. Figure 9 shows the surfaces of equal density  and the velocity field for subsonic and supersonic perturbers. The reflected wake overtakes the subsonic perturber after a time  $t=2d/(1-{\cal M})$.   For a perturber moving inside the medium parallel to its boundary, we choose ${\bm n}={\bm e}_x$ and ${\bm \xi}(t)= d{\bm e}_x+Vt{\bm e}_z$. The mirror trajectory ${\bm \xi}^r(t)= -d{\bm e}_x+Vt{\bm e}_z$ and the density perturbation and velocity potential are given as:
\begin{eqnarray}
\varrho_1&=& {\cal H}(x)\left[\varrho\left([(x-d)^2+y^2]^\frac{1}{2},z,V,t\right)+\varrho\left([(x+d)^2+y^2]^\frac{1}{2},z,V,t\right)\right],\\
\psi_1&=&{\cal H}(x)\left[\psi\left([(x-d)^2+y^2]^\frac{1}{2},z,V,t\right)+\psi\left([(x+d)^2+y^2]^\frac{1}{2},z,V,t\right)\right].
\end{eqnarray}
Figure 10 shows the corresponding surfaces and the velocity field for subsonic and supersonic perturbers in the plane $y=0$. Unlike motion normal to the boundary, the truncation of the wake is substantial. For $t\gg d/c$  the outgoing wake is reduced to approximately half its size in an infinite medium while the other half becomes the reflected wake. The presence of the two wakes on the same side of the perturber's trajectory  gives rise to a force normal to the direction of motion pushing the perturber away from the boundary. 

\newpage

\begin{figure}
\begin{center}
\includegraphics[width=75mm]{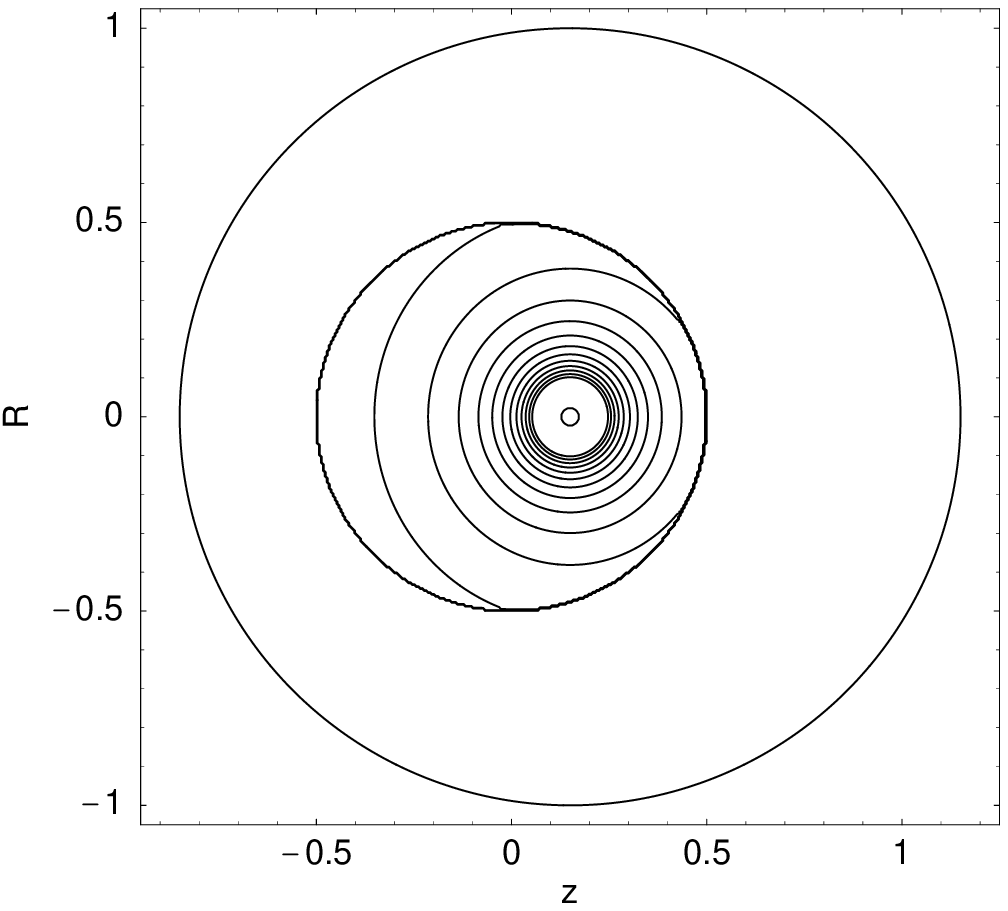}\includegraphics[width=75mm]{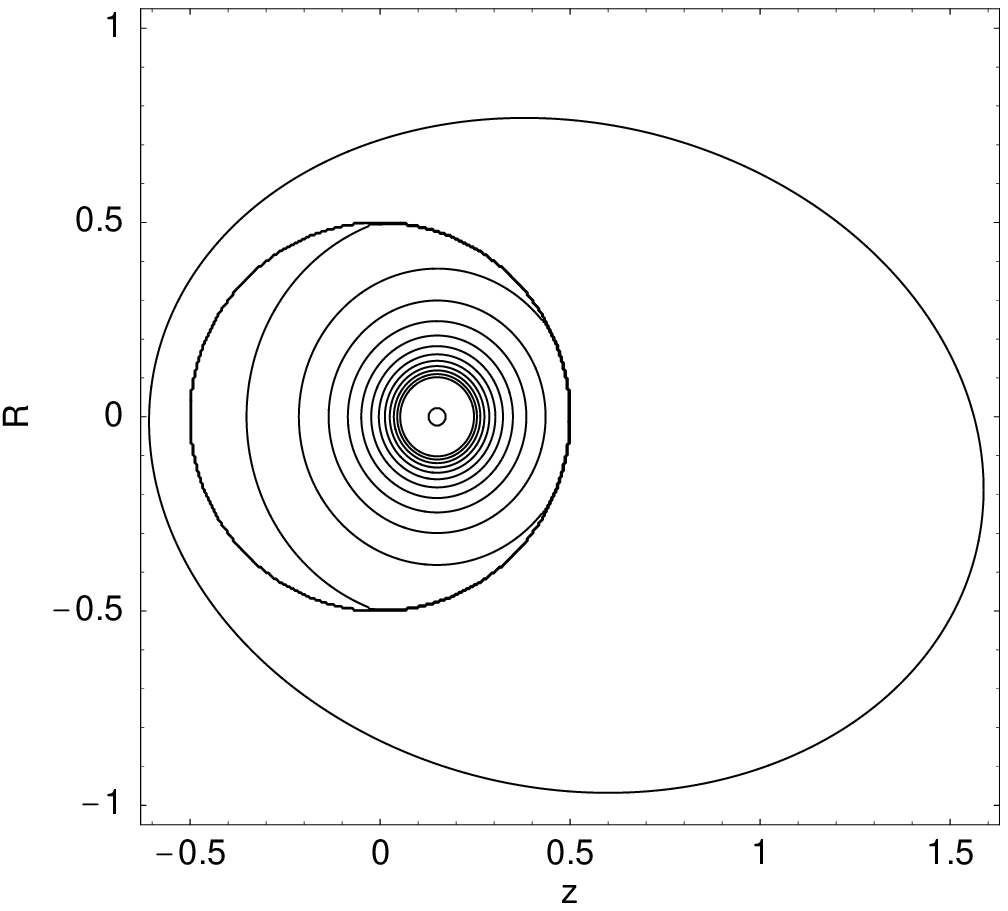}\\
\includegraphics[width=75mm]{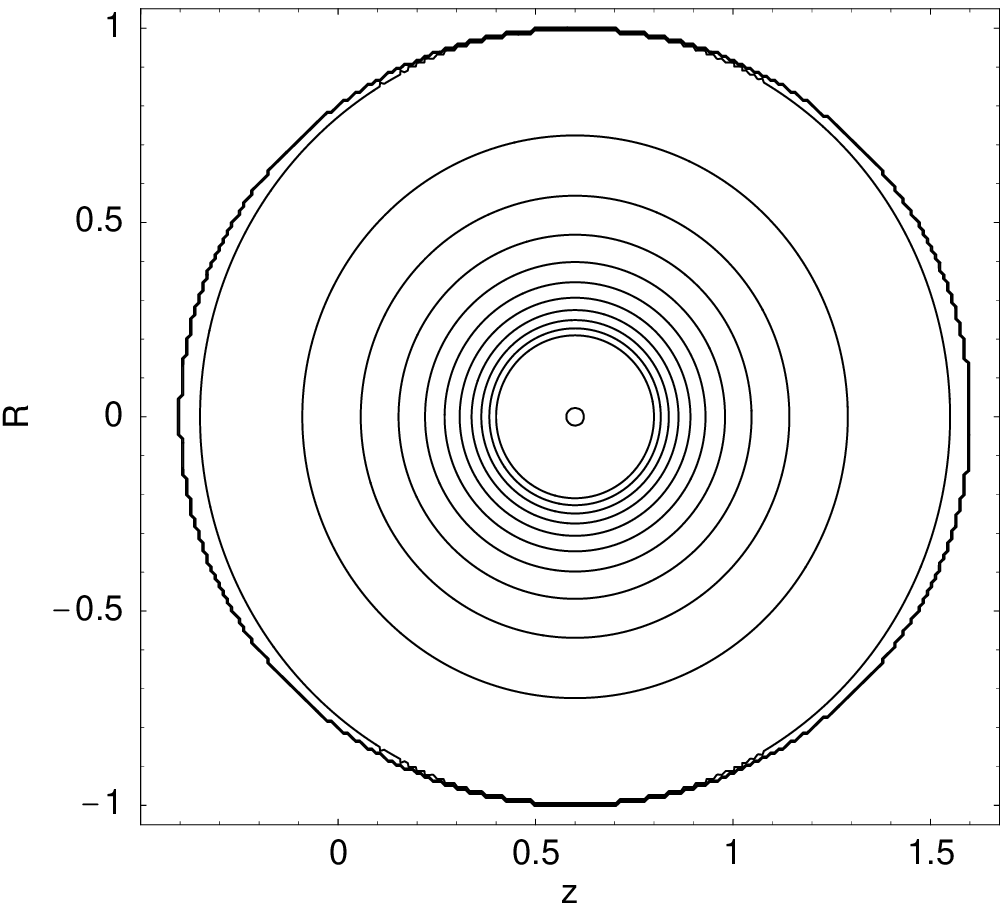}\includegraphics[width=75mm]{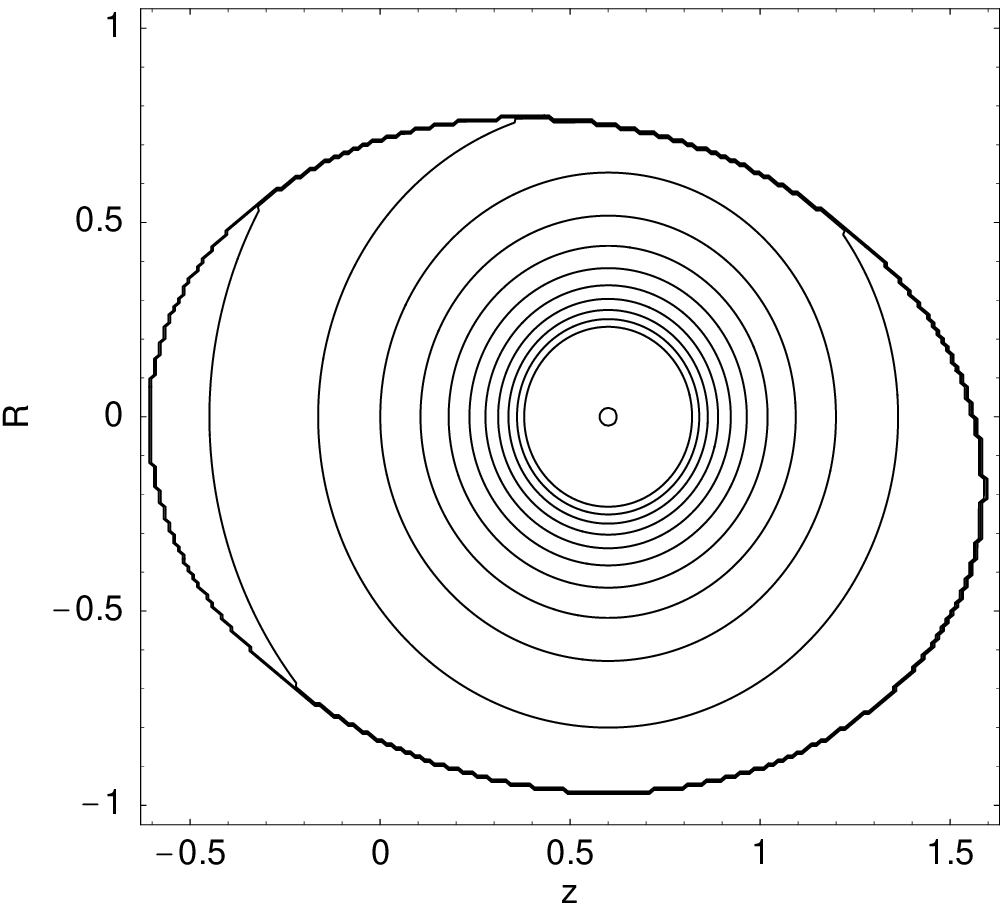}
\end{center}
\label{f1}
\caption{Surfaces of equal density (\ref{density2}) for a subsonic perturber (${\cal M}=0.3$) in a medium with a comoving symmetric boundary (left column) and a medium with a boundary unrelated to the perturber (right column). Distances are scaled to $\rmx$. At $t=\rmx/2c$, the sonic wavefront (bold line) launched at $t=0$ has not reached  the boundary (top row) and the forces in the two media are identical and given by (\ref{force3},\ref{ostriker}). In the steady state regime, $t\geq 2\rmx/c$ (bottom row), the density perturbation is symmetric with respect to the perturber for the medium with the comoving symmetric boundary and the perturber is not subject to friction (bottom left panel) (equations \ref{force3},\ref{steady1}). The boundary of the stationary medium (bottom right panel) truncates the density perturbation asymmetrically and results in finite force on the subsonic perturber.}
\end{figure}

\newpage

\begin{figure}
\begin{center}
\includegraphics[width=75mm]{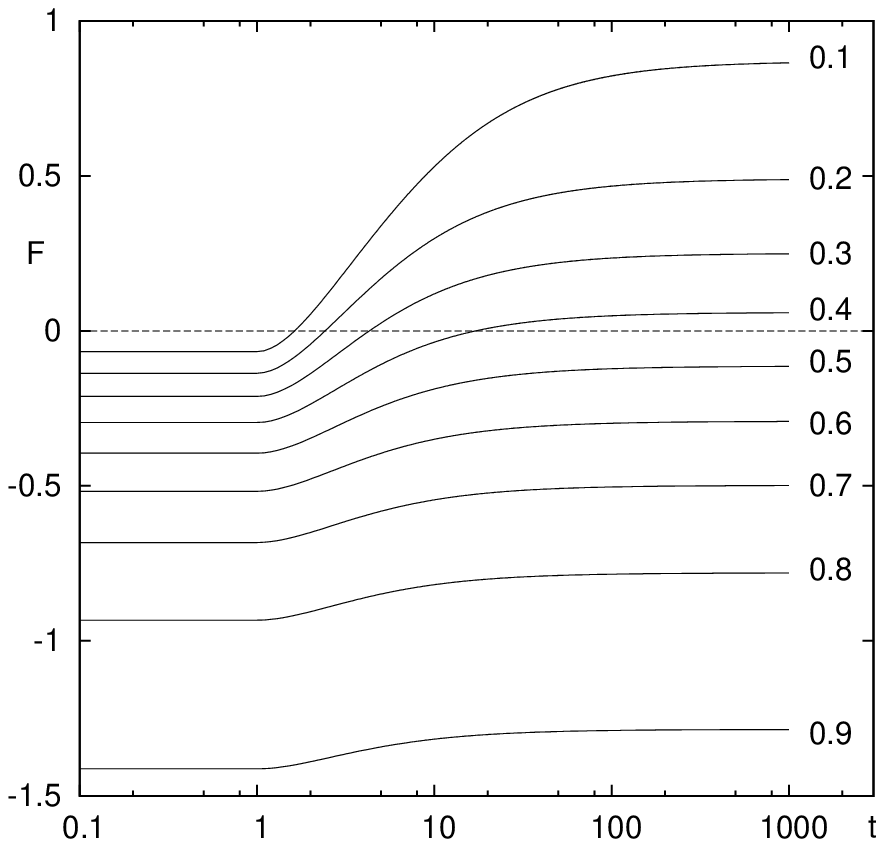}\includegraphics[width=75mm]{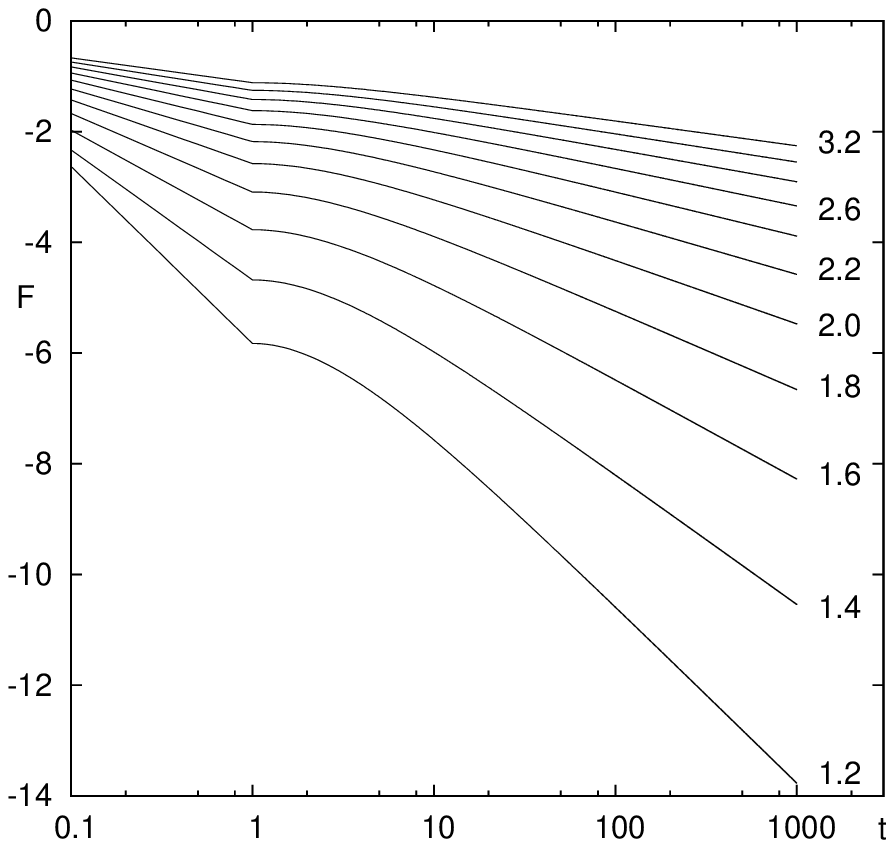}\\
\includegraphics[width=75mm]{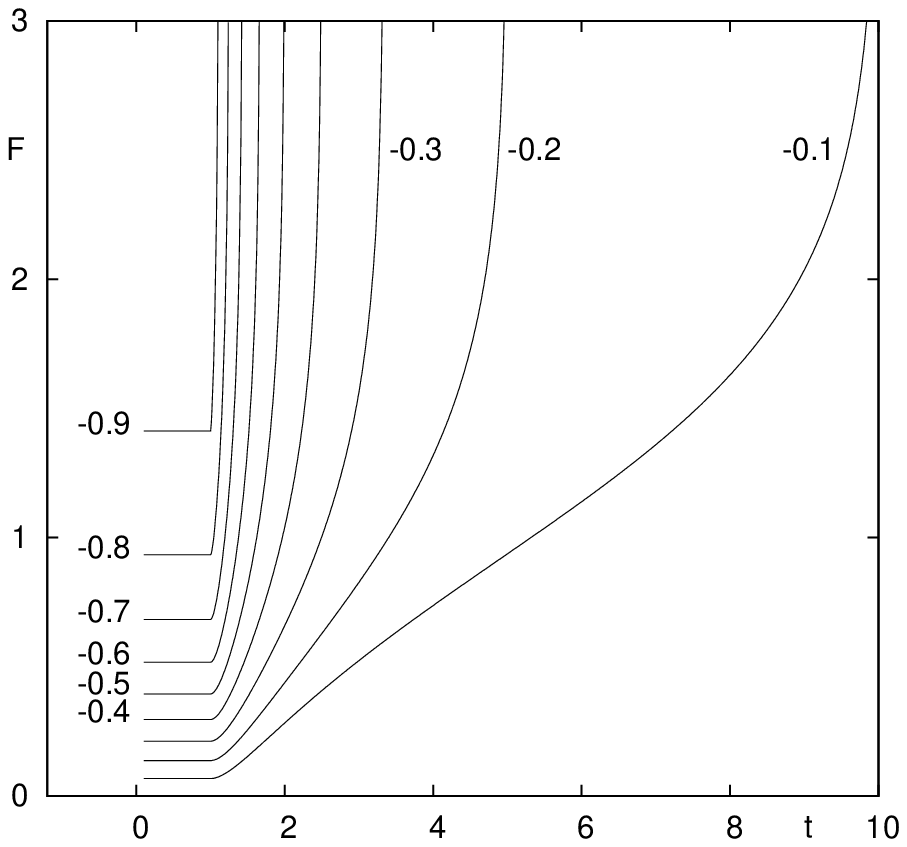}\includegraphics[width=75mm]{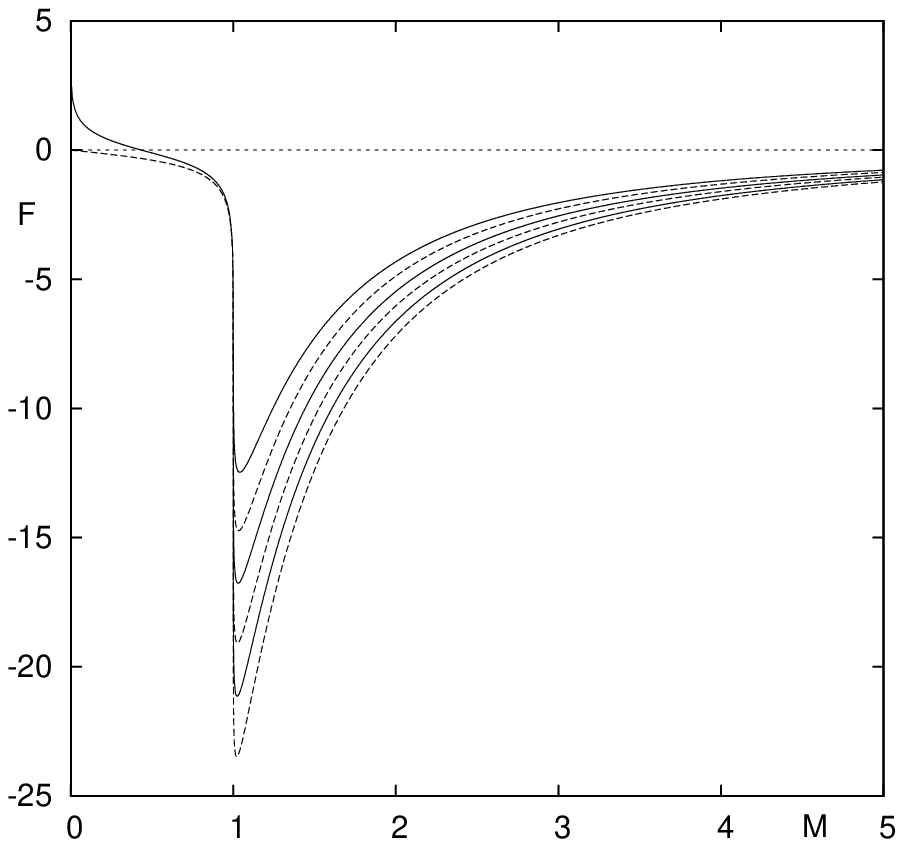}
\end{center}
\label{f2}
\caption{Friction force from the outgoing wake for motion normal to the boundary. The force scaled by $2\pi(GM)^2\rho_0/c^2$ is given by $I/{\cal M}^2$ (\ref{incidentsub},\ref{incidentsuper}); time is scaled by $d/c$. The force is shown as a function of time for subsonic perturbers, 
${\cal M}=0$ to $0.9$ (top left), supersonic perturbers, ${\cal M}=1.2$ to $3.2$, (top right) and for subsonic motion towards the boundary ${\cal M}=-0.9$ to $-0.1$ (bottom left). The limit force as $t\gg d/c$ is shown a function of the Mach number (bottom right) along the force experienced in an infinite medium (\ref{ostriker}) (dotted line) for $r_0= 10^{-2}\, d$ and $ct/d=10^2, \ 10^3$ and  $10^4.$}
\end{figure}

\begin{figure}
\begin{center}
\includegraphics[width=75mm]{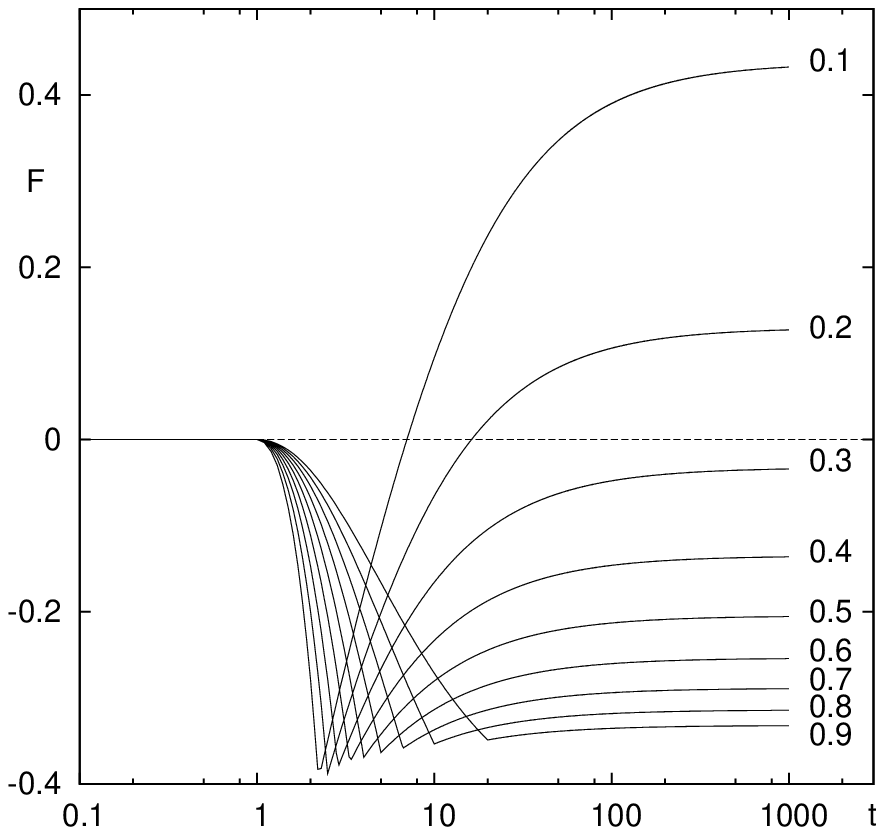}\includegraphics[width=75mm]{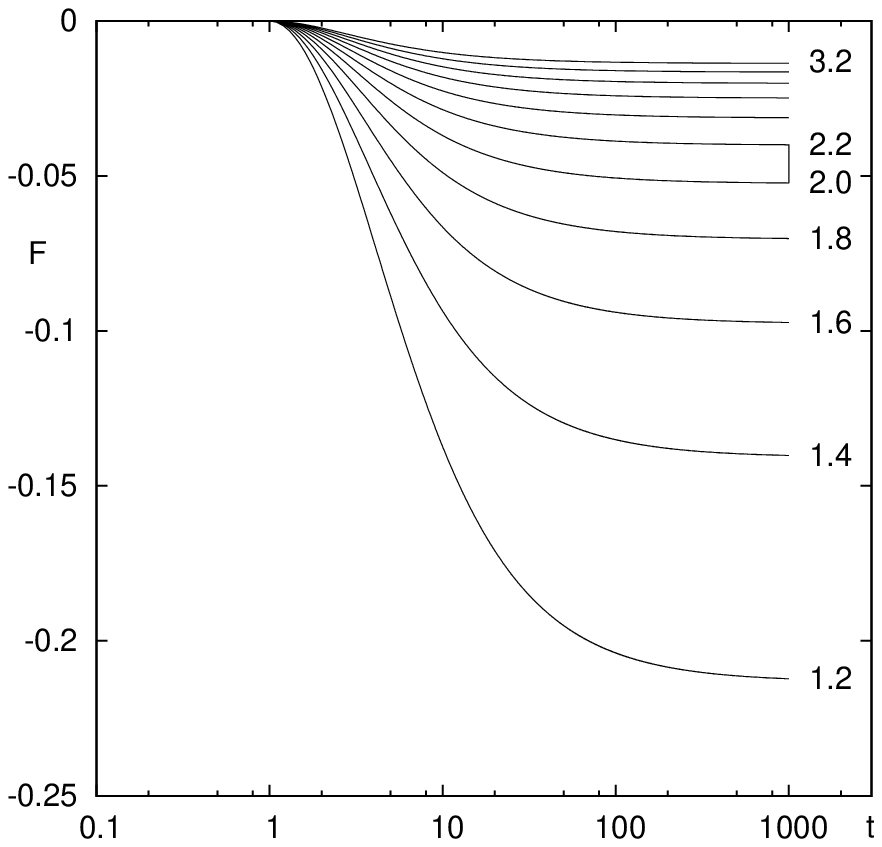}\\
\includegraphics[width=75mm]{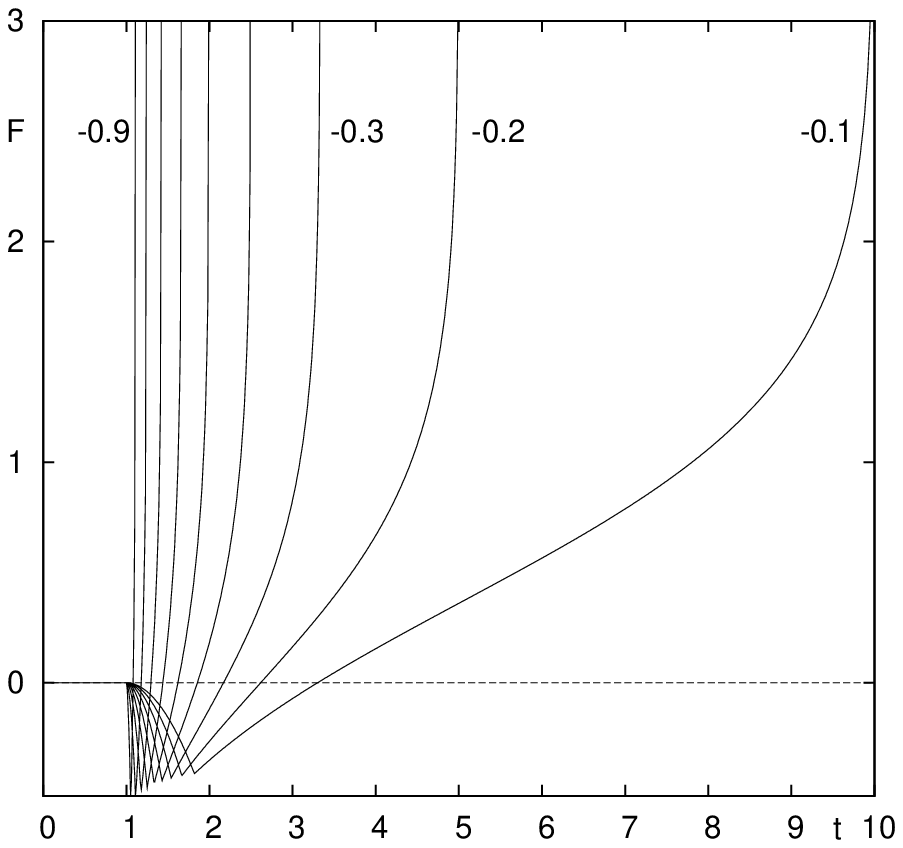}\includegraphics[width=75mm]{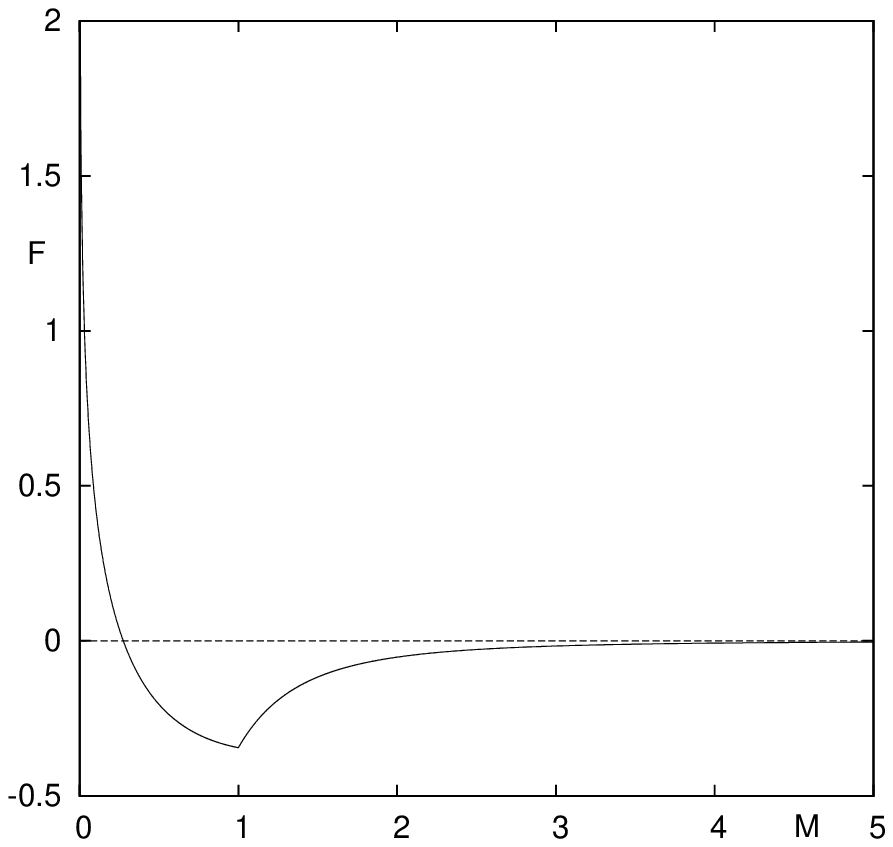}
\end{center}
\label{f3}
\caption{Friction force from the reflected wake for motion normal to the boundary. The plots' parameters are identical to those in Figure 2.}
\end{figure}

\begin{figure}
\begin{center}
\includegraphics[width=75mm]{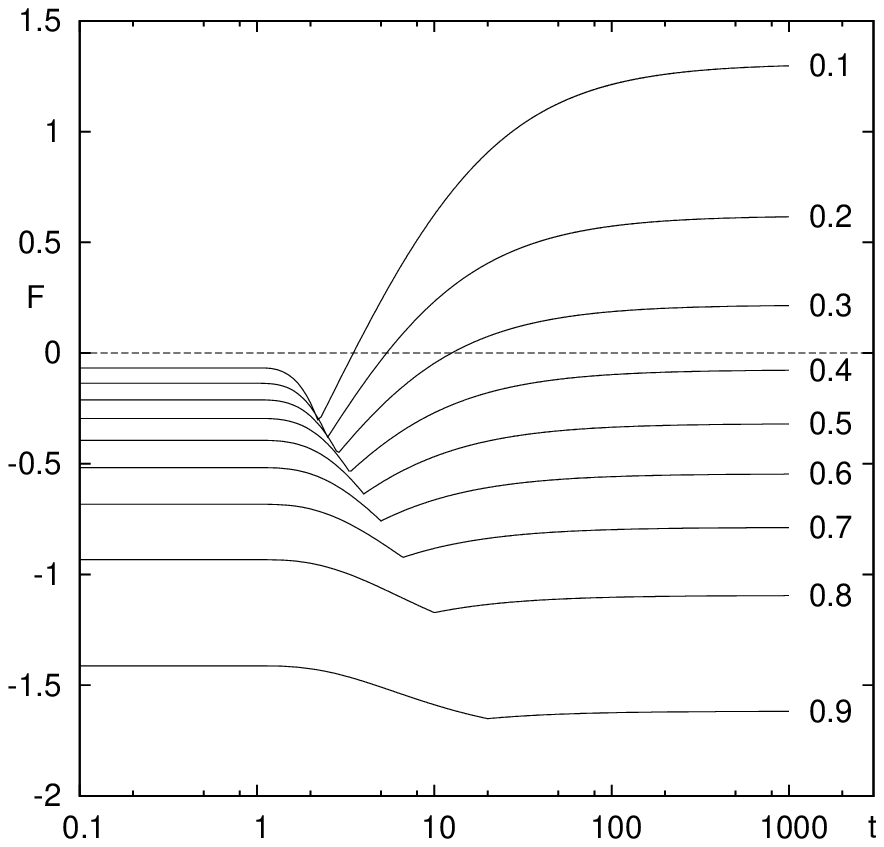}\includegraphics[width=75mm]{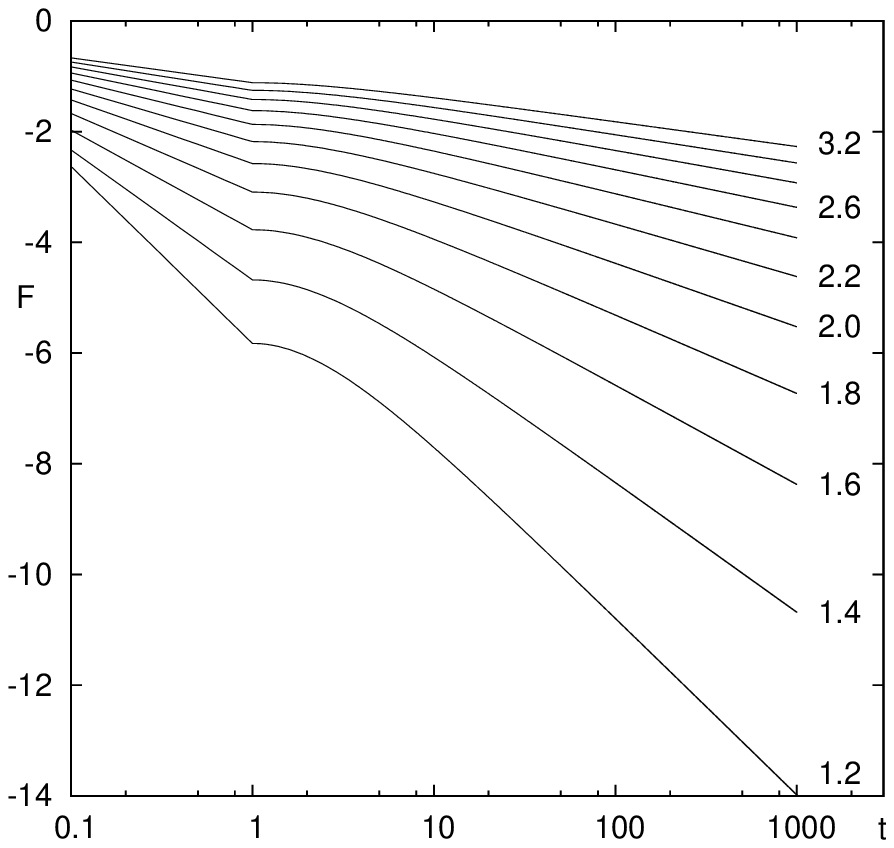}\\
\includegraphics[width=75mm]{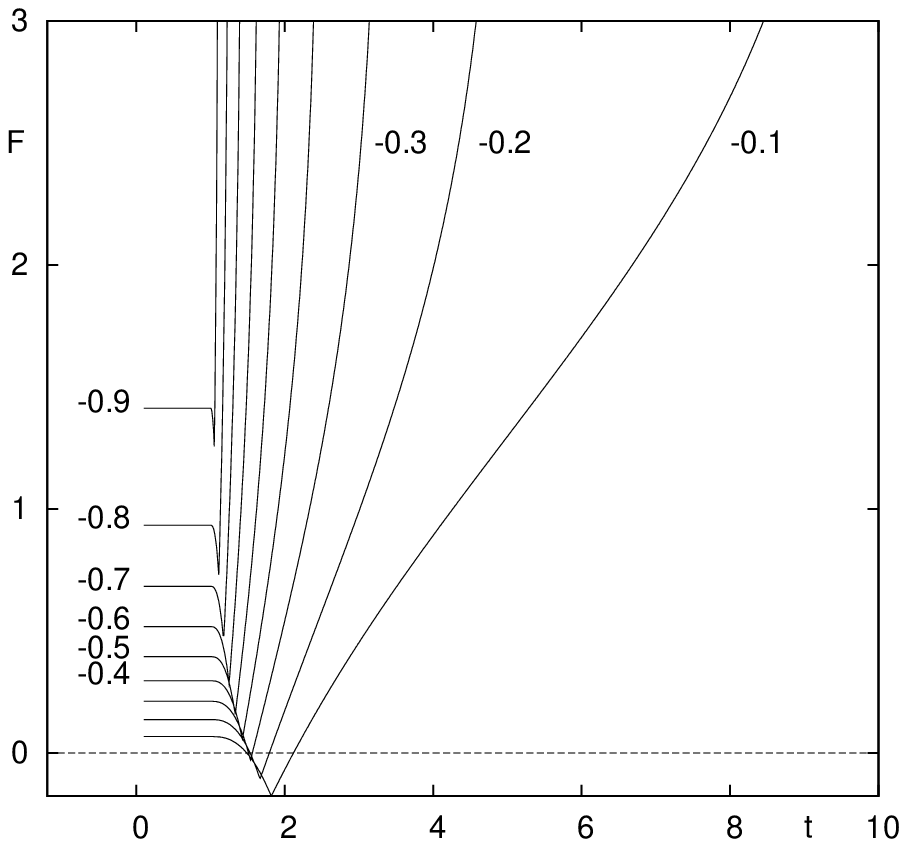}\includegraphics[width=75mm]{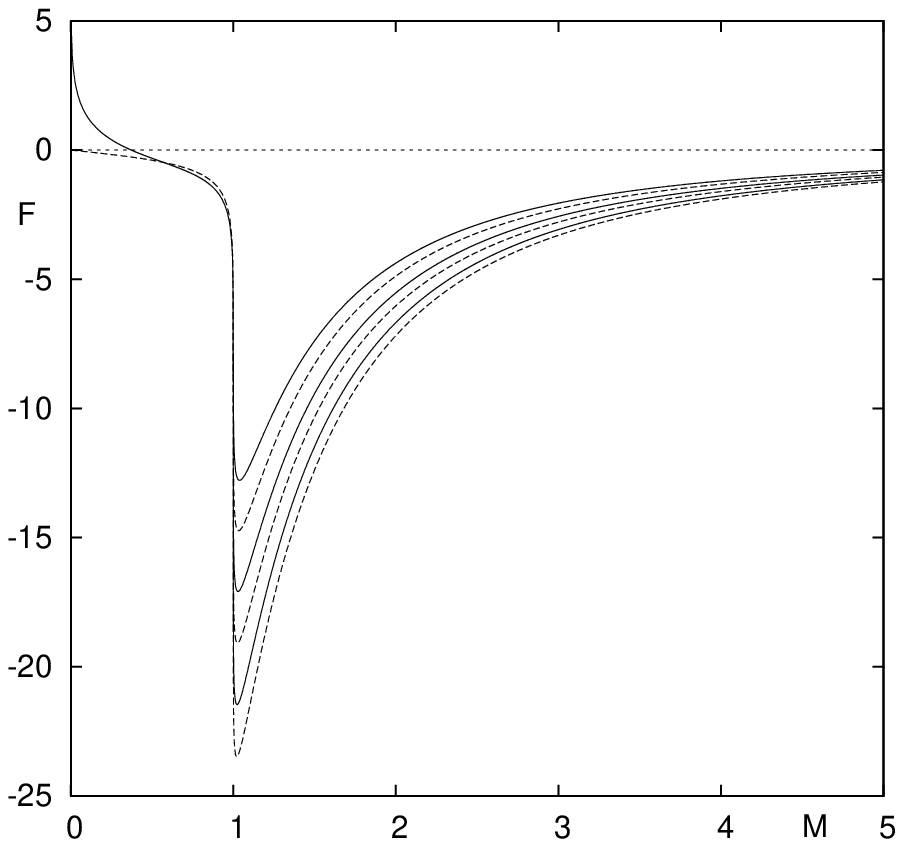}
\end{center}
\label{f4}
\caption{Total friction force from outgoing and reflected wakes for motion normal to the boundary. The plots' parameters are identical to those in Figure 2.}
\end{figure}

\begin{figure}
\begin{center}
~\\[-24mm]
\includegraphics[width=72mm]{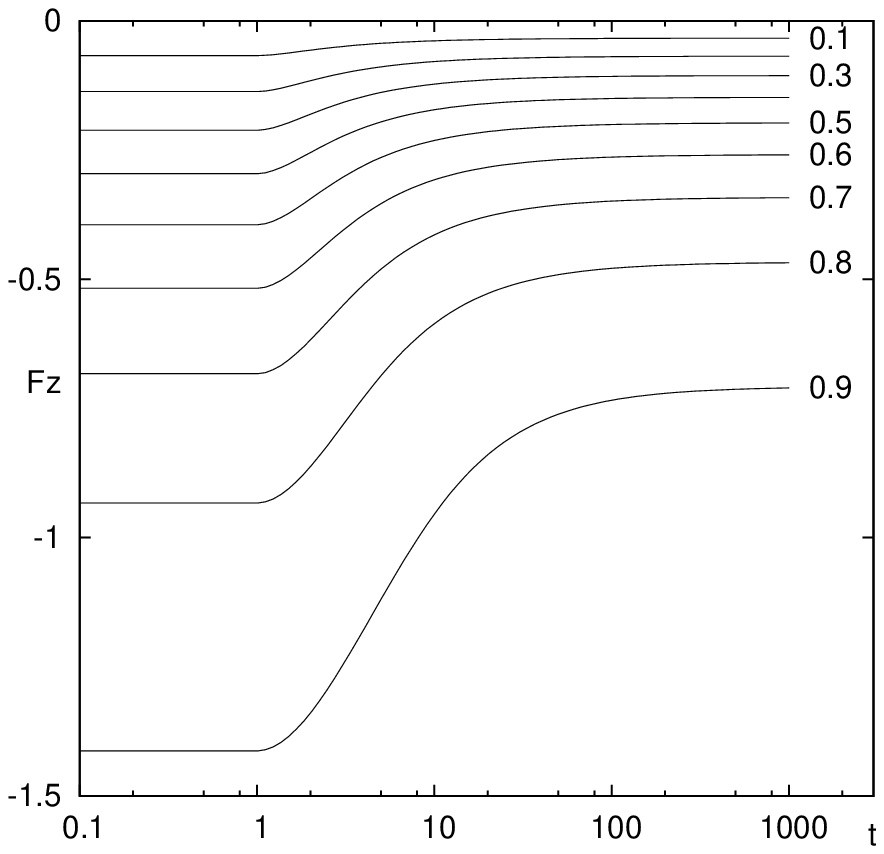}\includegraphics[width=72mm]{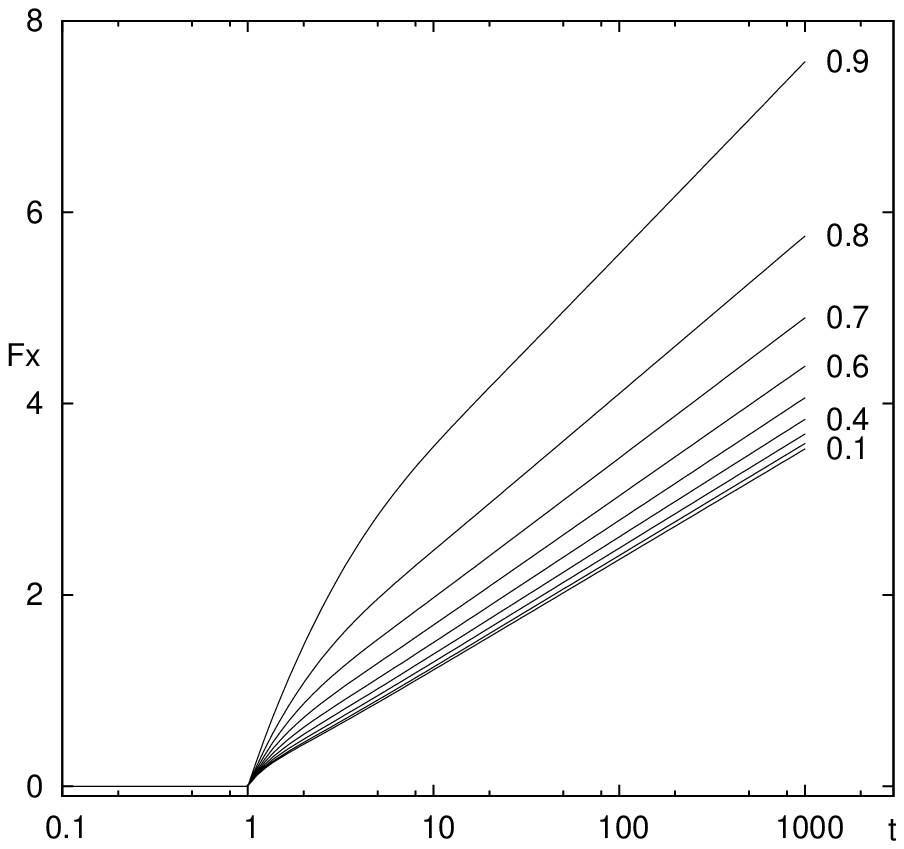}\\
\includegraphics[width=72mm]{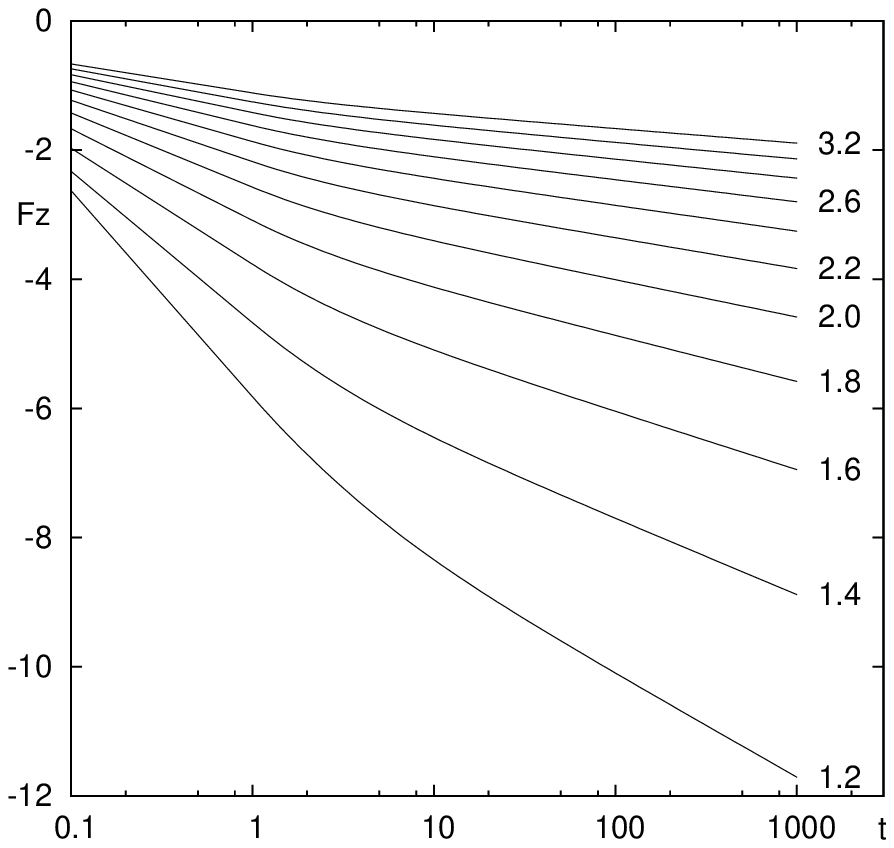}\includegraphics[width=72mm]{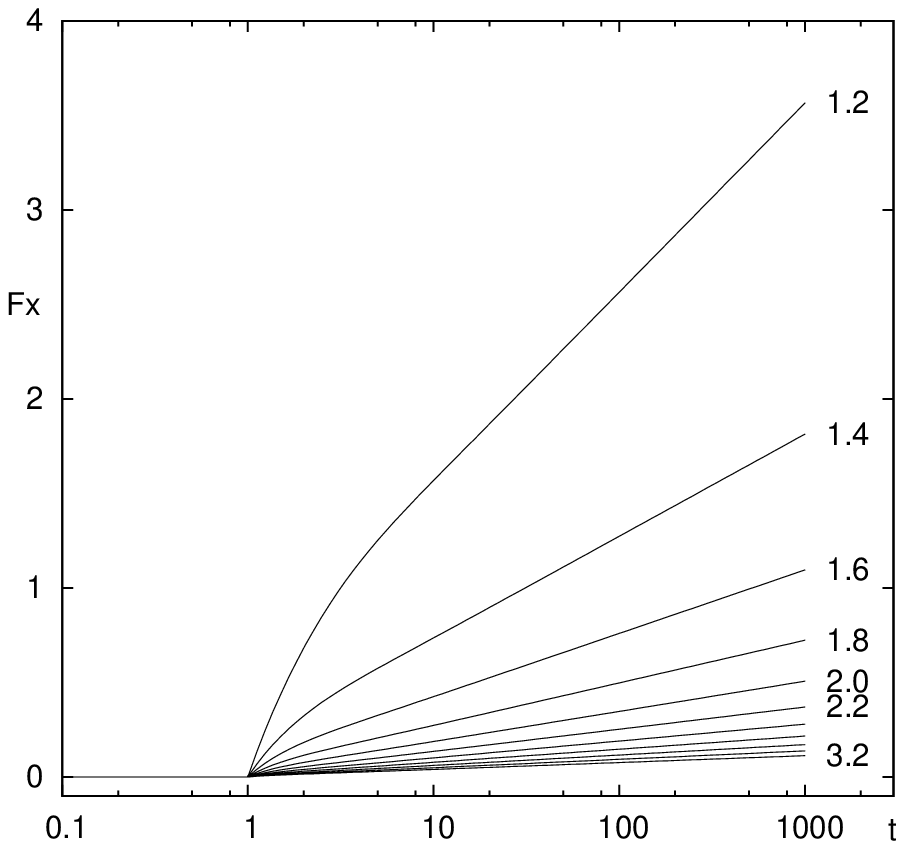}\\
\includegraphics[width=72mm]{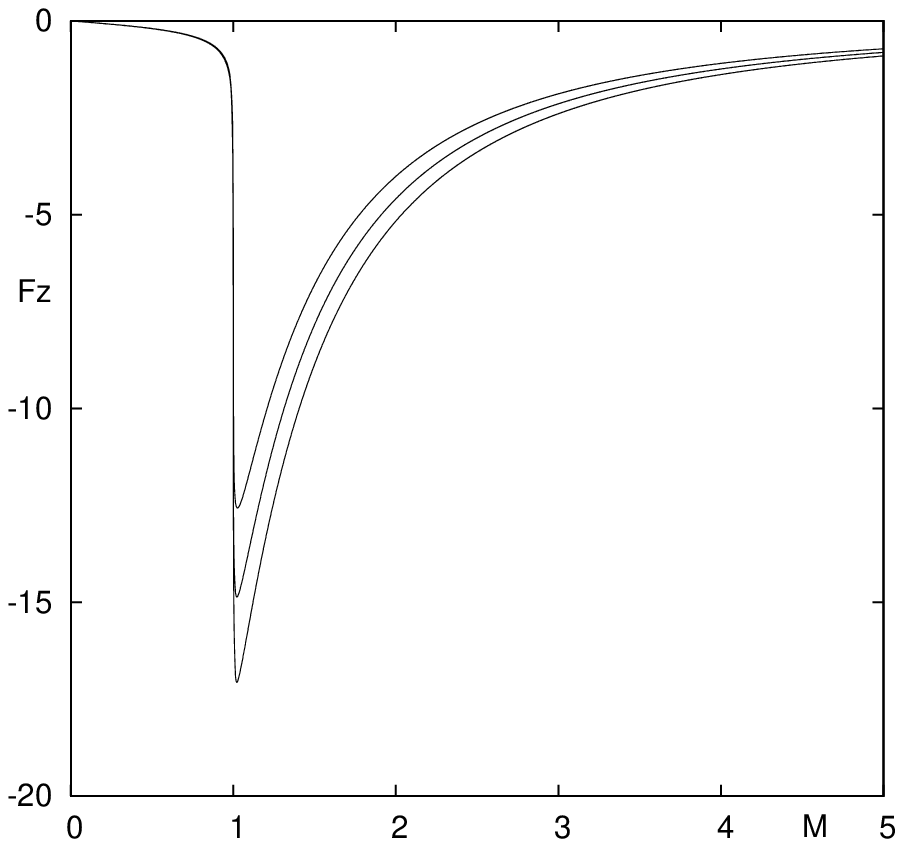}\includegraphics[width=72mm]{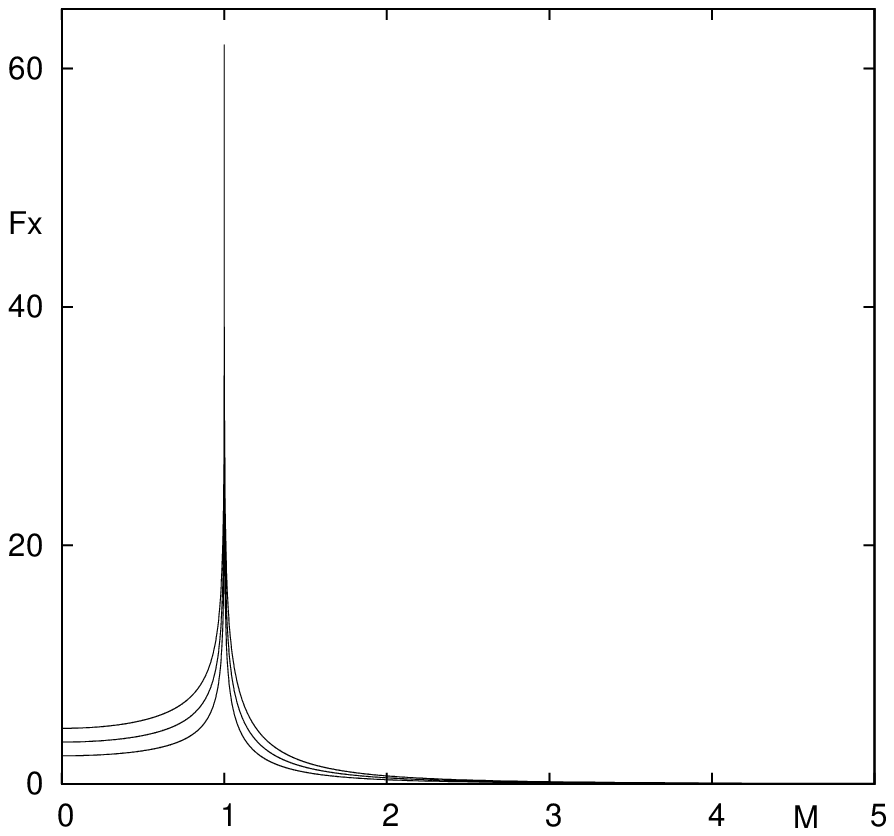}
\end{center}
\label{f5}
\vspace*{-12mm}
\caption{Friction force from the outgoing wake for motion parallel to the boundary. The force scaled by $2\pi(GM)^2\rho_0/c^2$. The component along the trajectory (left column) is given by $F_z(r_0/|1+{\cal M}|\leq r\leq d)+F_z(d\leq r \leq ct)$ where the first term is obtained from (\ref{ostriker}) and the second term from (\ref{forcepz2}). The component normal to the trajectory (right column) is $F_x(d\leq r \leq ct)$ given by (\ref{forcepx2}); time is scaled by $d/c$. The force is shown as a function of time for subsonic perturbers, 
${\cal M}=0$ to $0.9$ (top row), supersonic perturbers, ${\cal M}=1.2$ to $3.2$, (middle row). The limit force as $t\gg d/c$ is shown a function of the Mach number (bottom row) for $r_0= 10^{-2}\, d$ and $ct/d=10^2, \ 10^3$ and  $10^4.$}
\end{figure}

\begin{figure}
\begin{center}
\includegraphics[width=75mm]{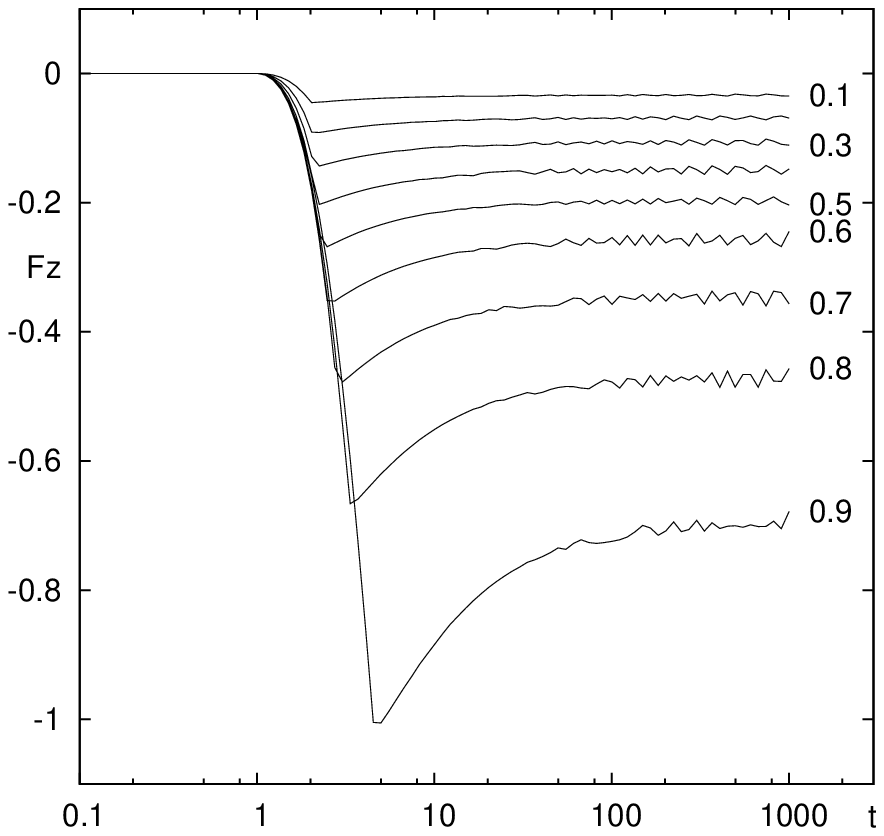}\includegraphics[width=75mm]{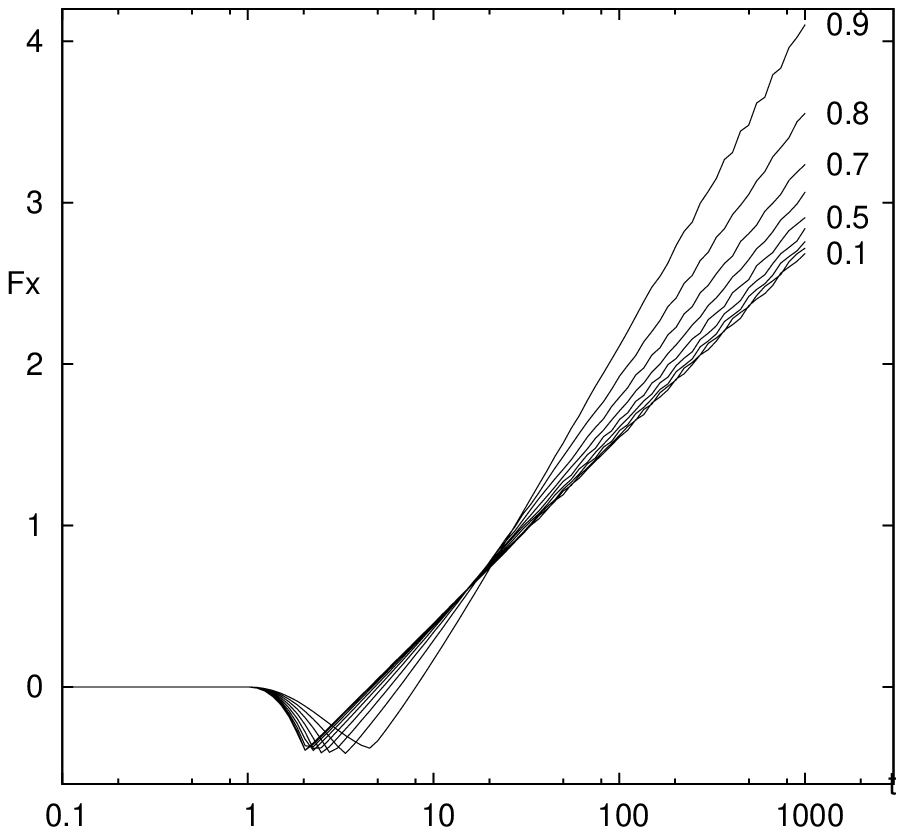}\\
\includegraphics[width=75mm]{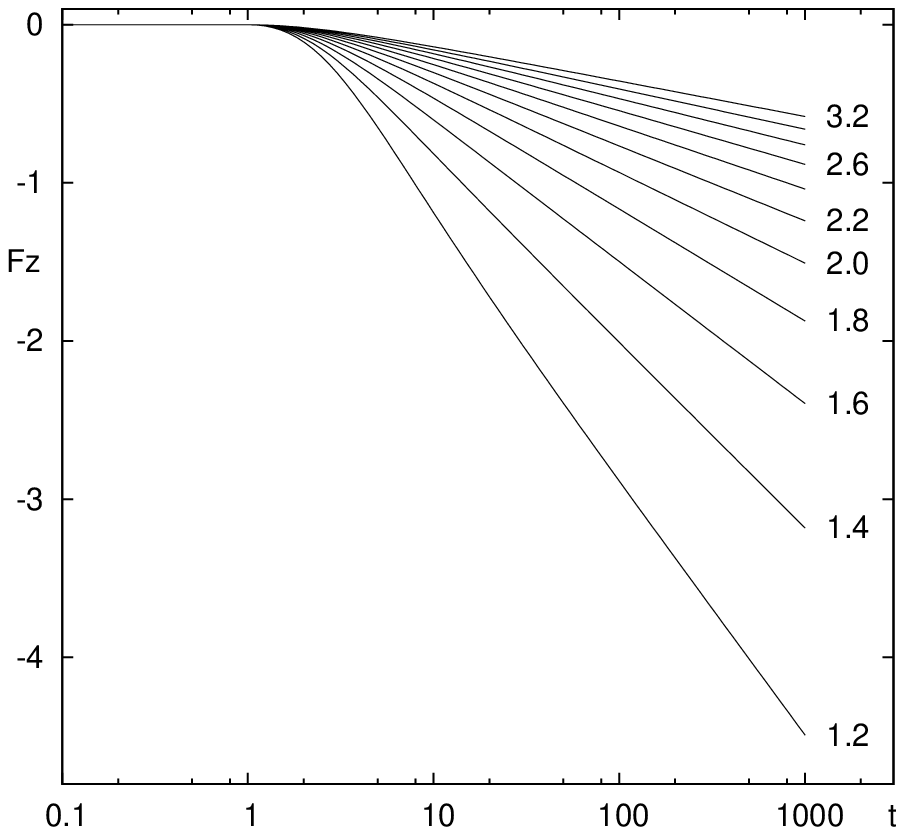}\includegraphics[width=75mm]{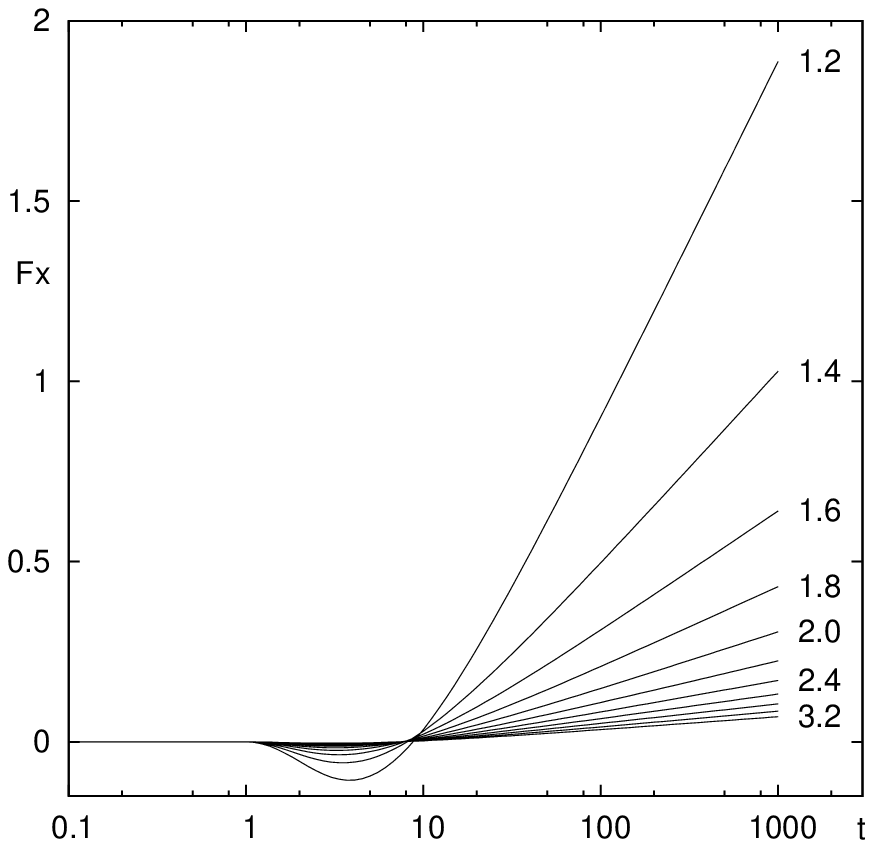}\\
\includegraphics[width=75mm]{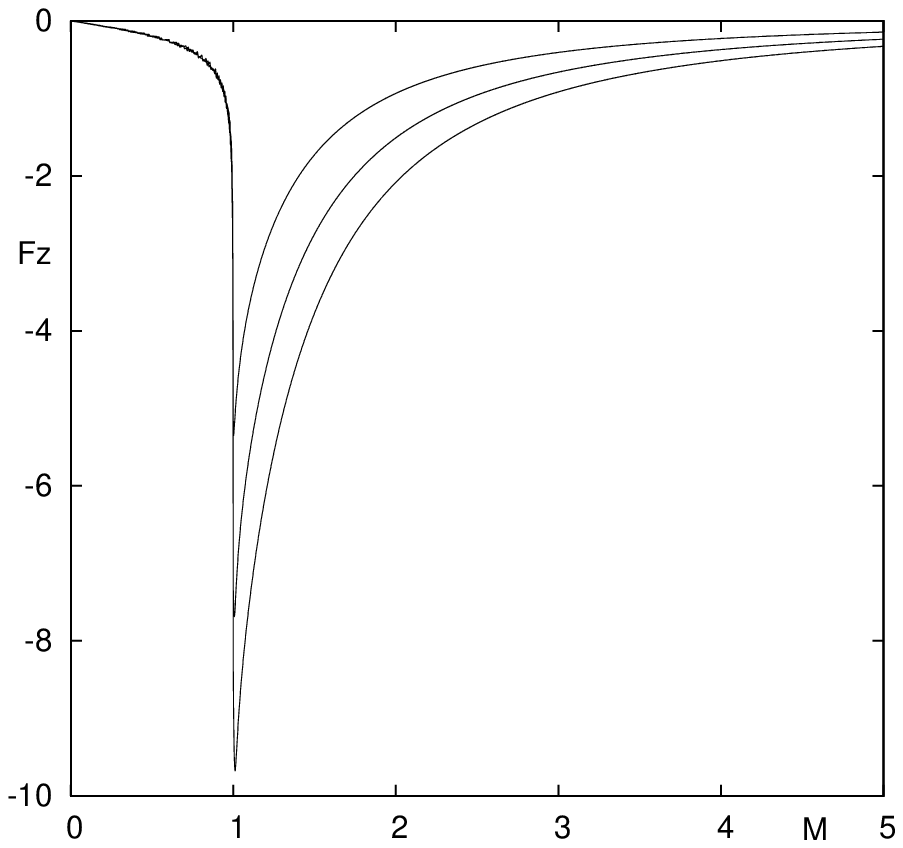}\includegraphics[width=75mm]{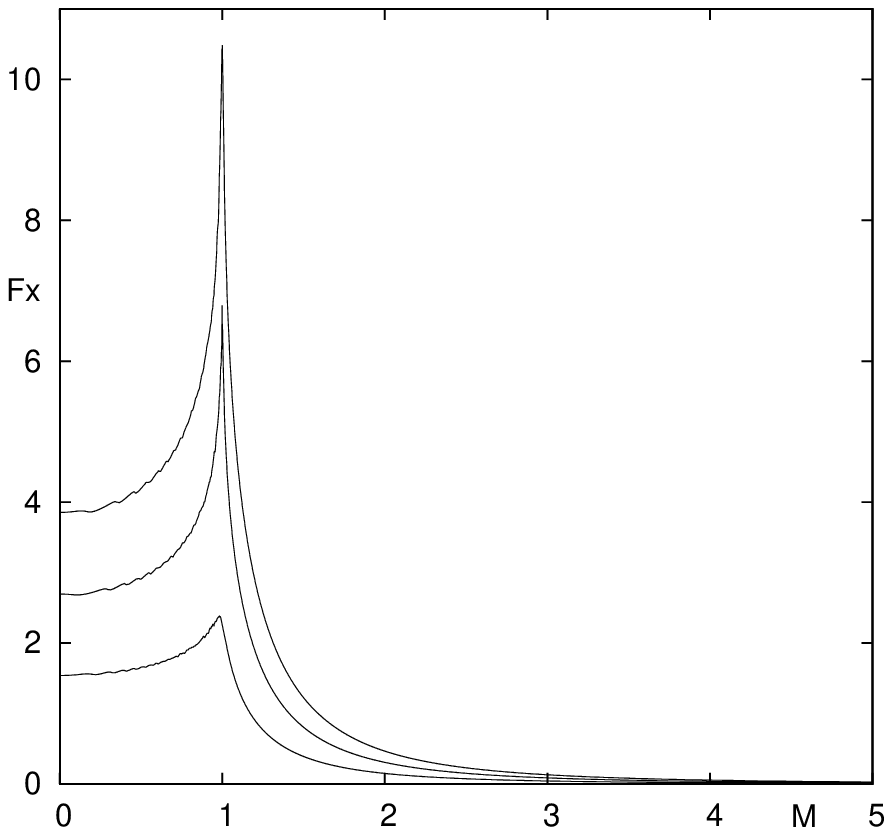}
\end{center}
\label{f6}
\caption{Friction force from the reflected wake for motion parallel to the boundary. The plots' parameters are identical to those in Figure 5.}
\end{figure}

\begin{figure}
\begin{center}
~\\[-12mm]
\includegraphics[width=75mm]{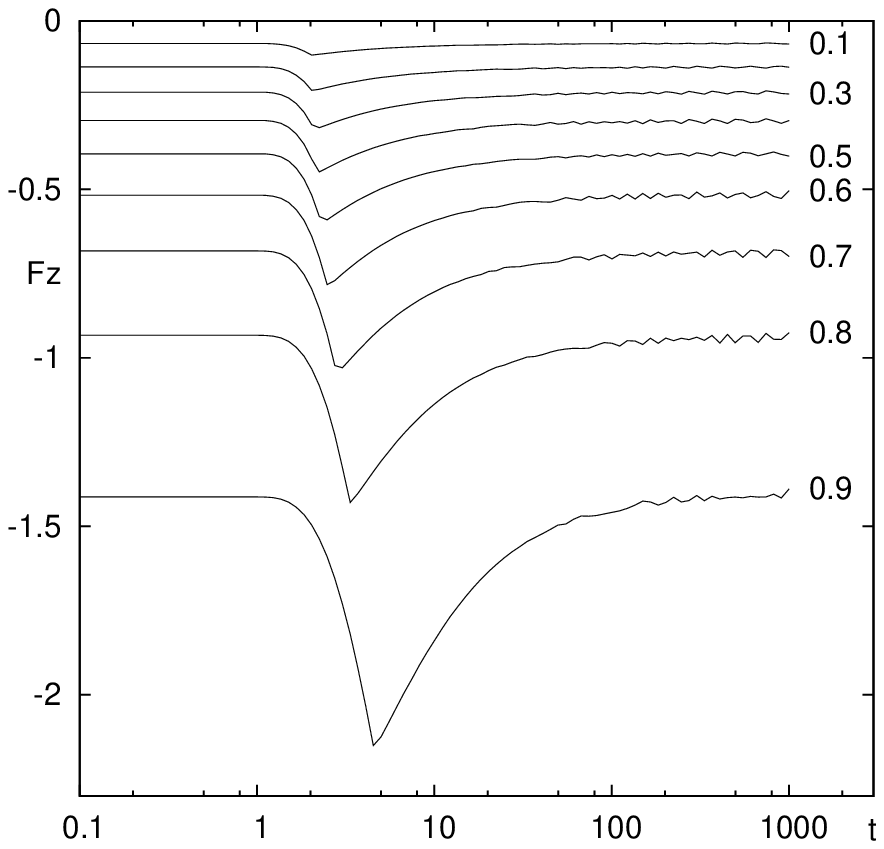}\includegraphics[width=75mm]{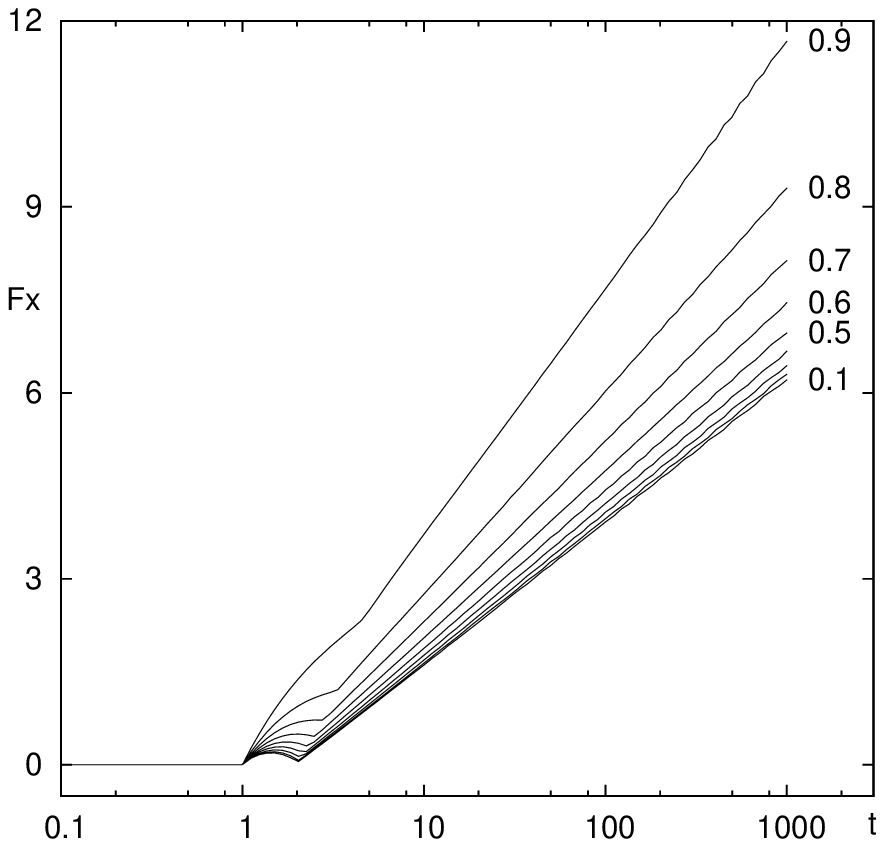}\\
\includegraphics[width=75mm]{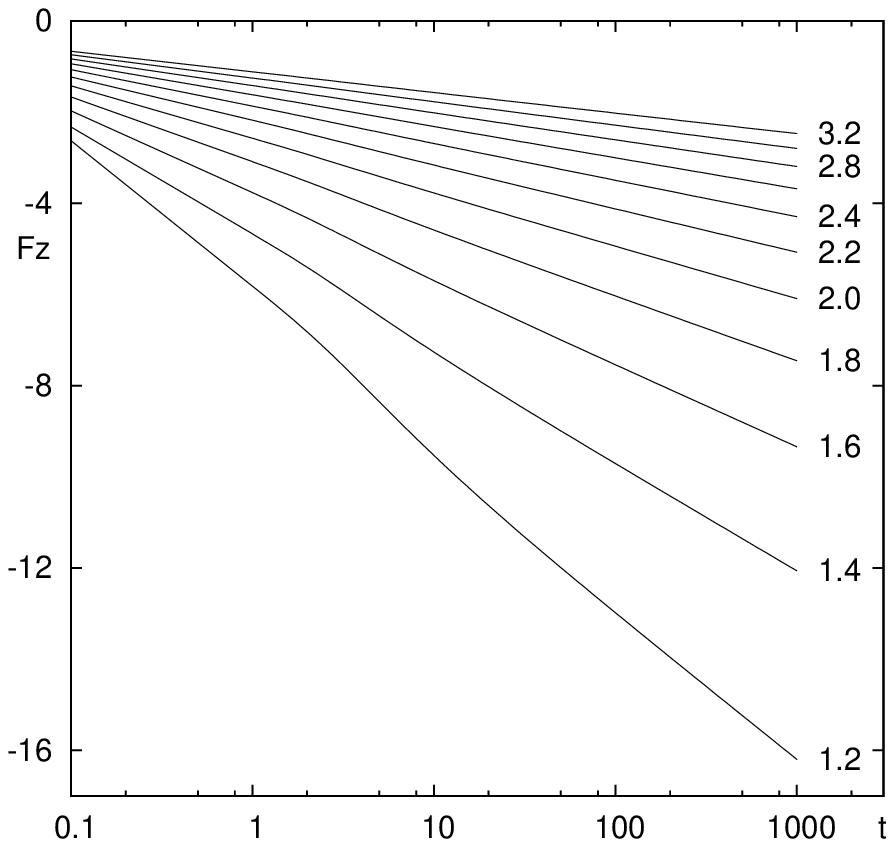}\includegraphics[width=75mm]{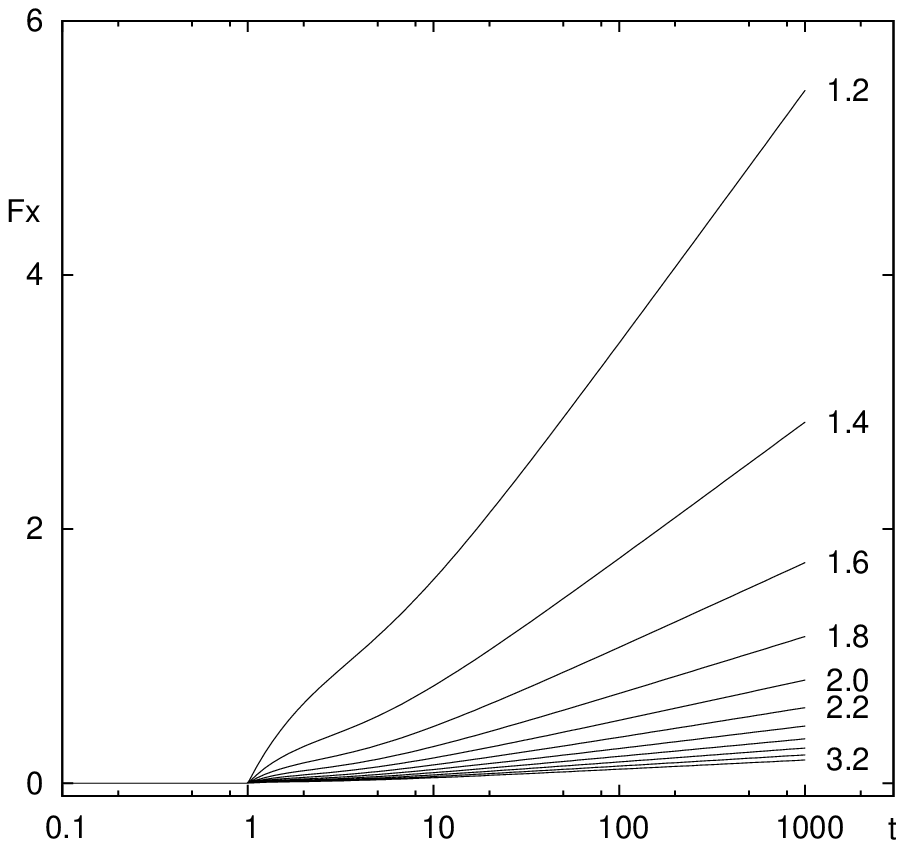}\\
\includegraphics[width=75mm]{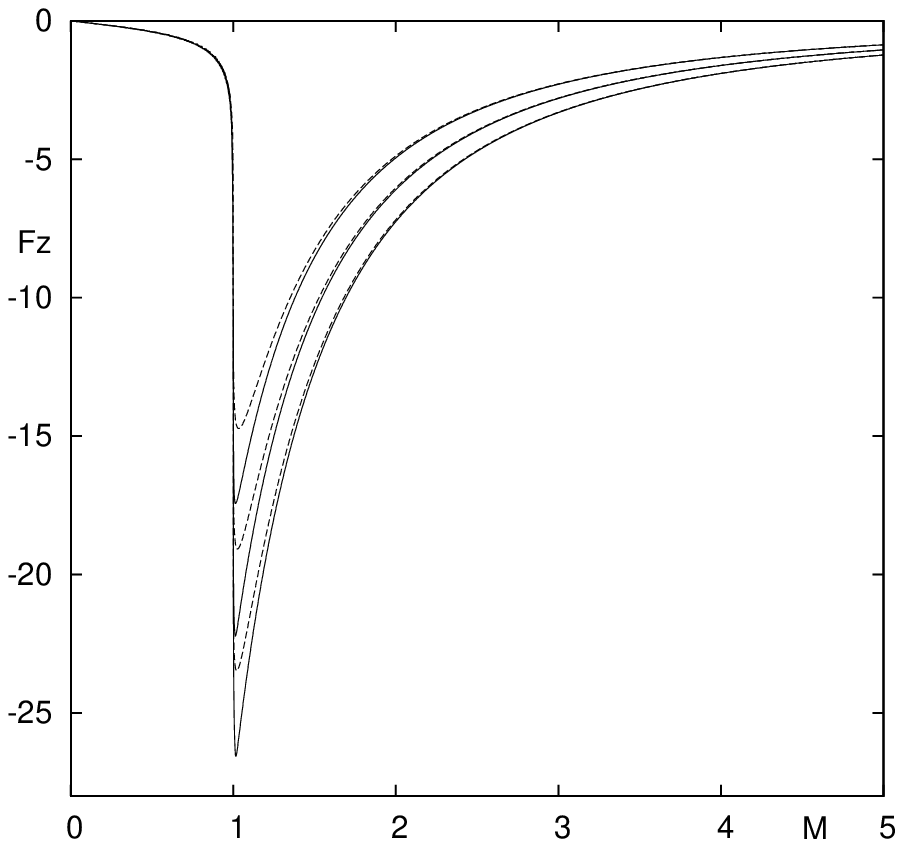}\includegraphics[width=75mm]{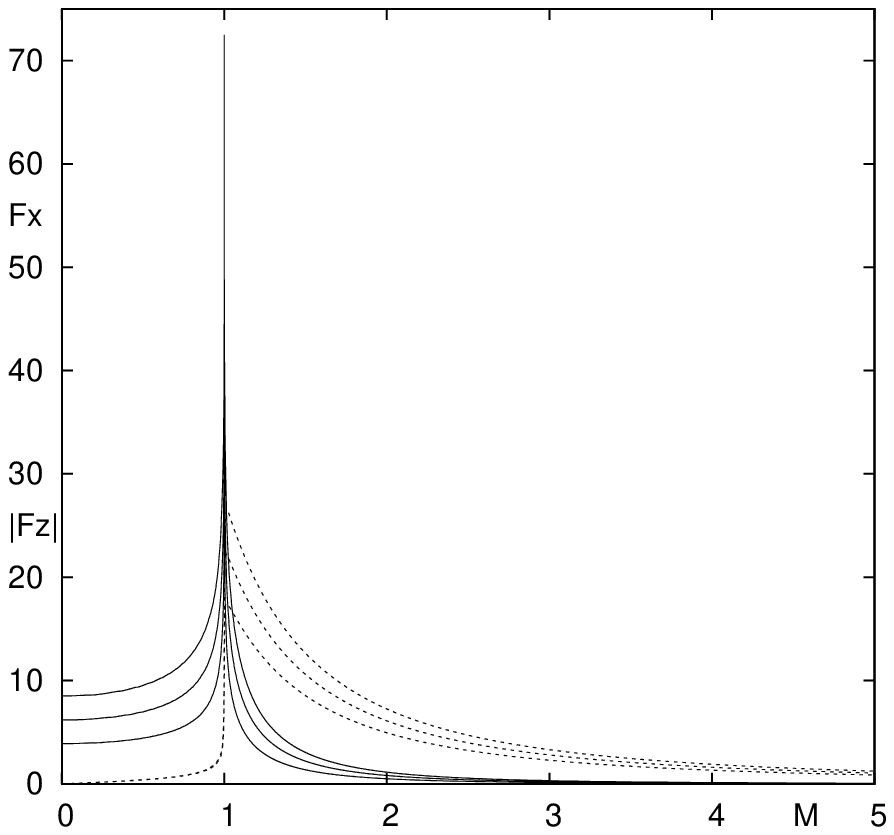}
\end{center}
\vspace*{-12mm}
\label{f7}
\caption{Friction force from the outgoing and reflected  
wakes for motion parallel to the boundary. The plots' parameters are identical to those in Figure 5. In the bottom left panel, the force along the trajectory for an infinite medium (\ref{ostriker}) is shown for comparison (dashed line). In the bottom right panel, both $F_x$ (solid line) and $|F_z|$ (dashed line) are shown as functions of the Mach number.}
\end{figure}

\begin{figure}
\begin{center}
\ \ \ \ \includegraphics[width=70mm]{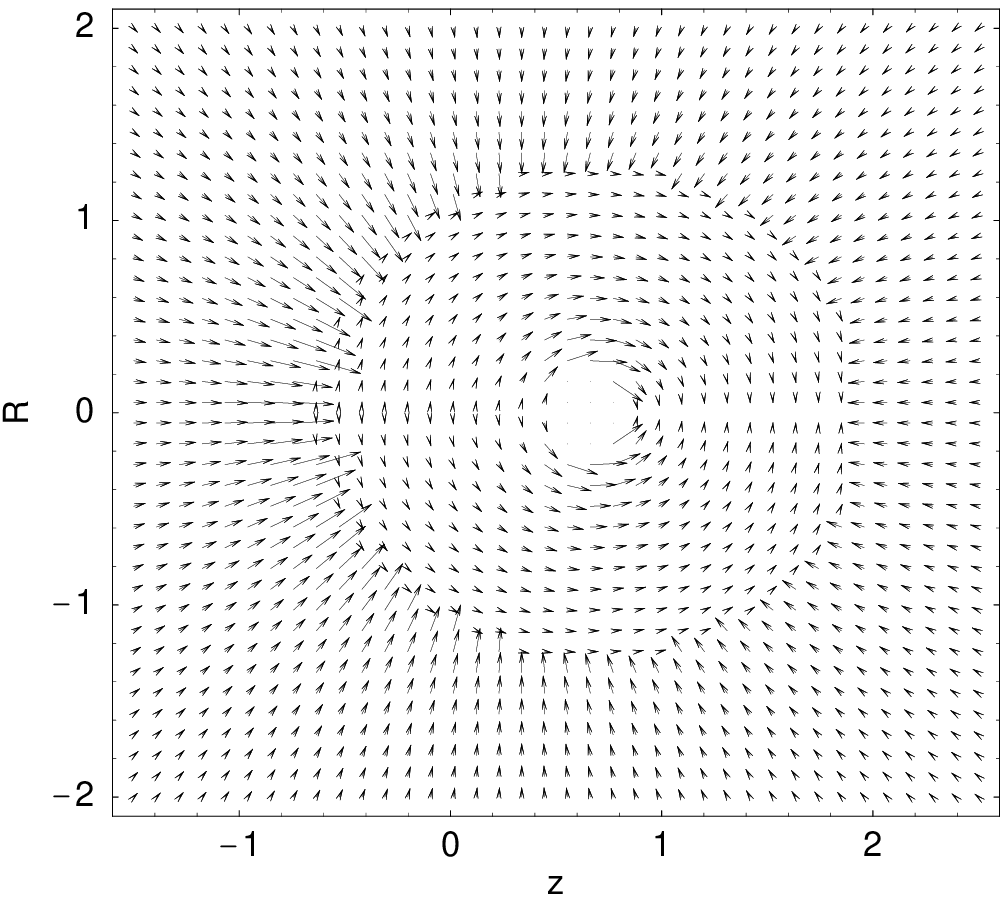}\includegraphics[width=73mm]{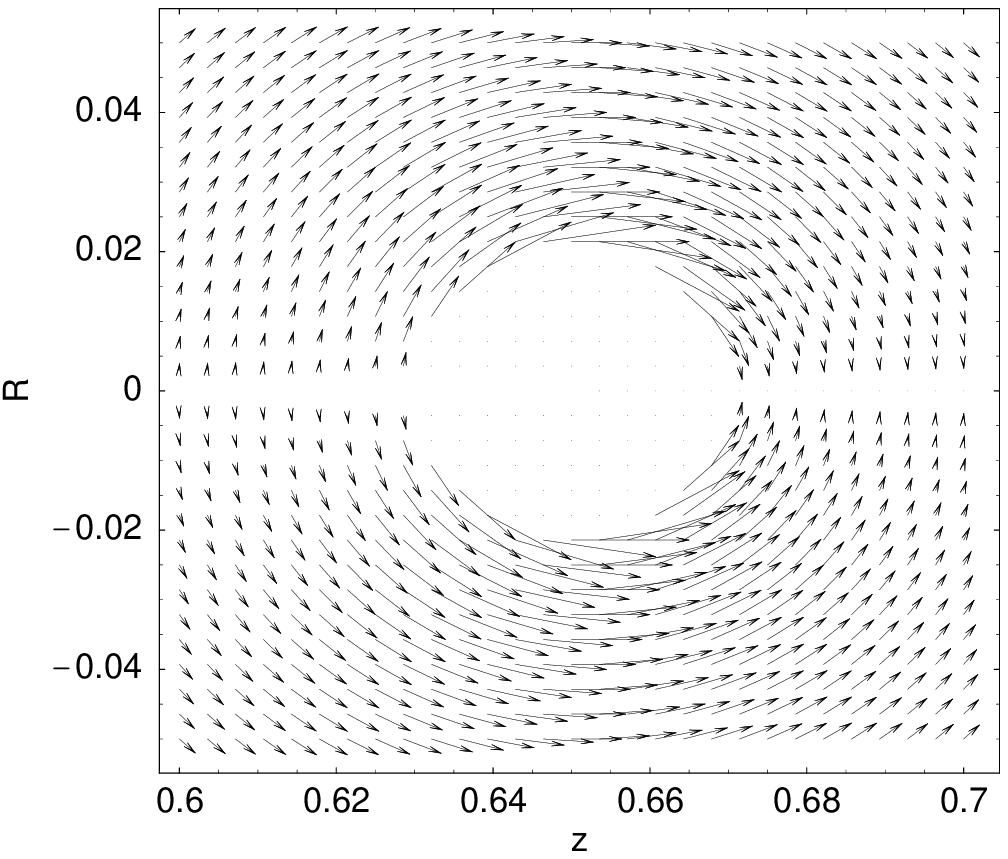}\\
\includegraphics[width=75.5mm]{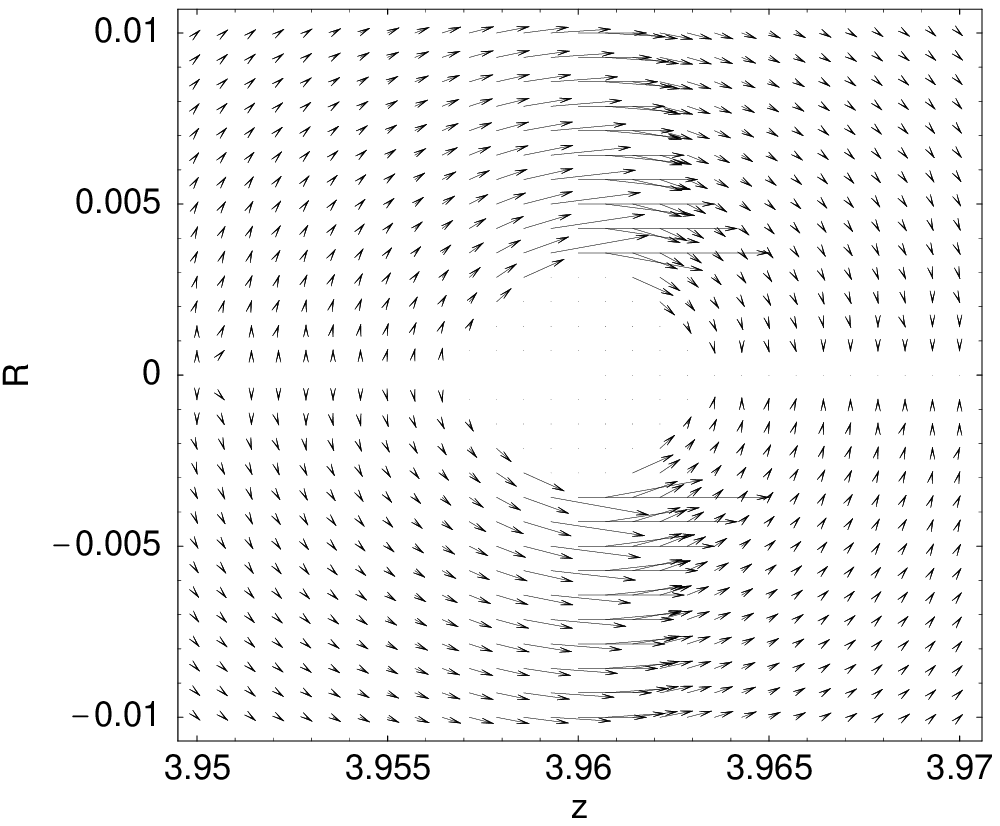}\includegraphics[width=72.5mm]{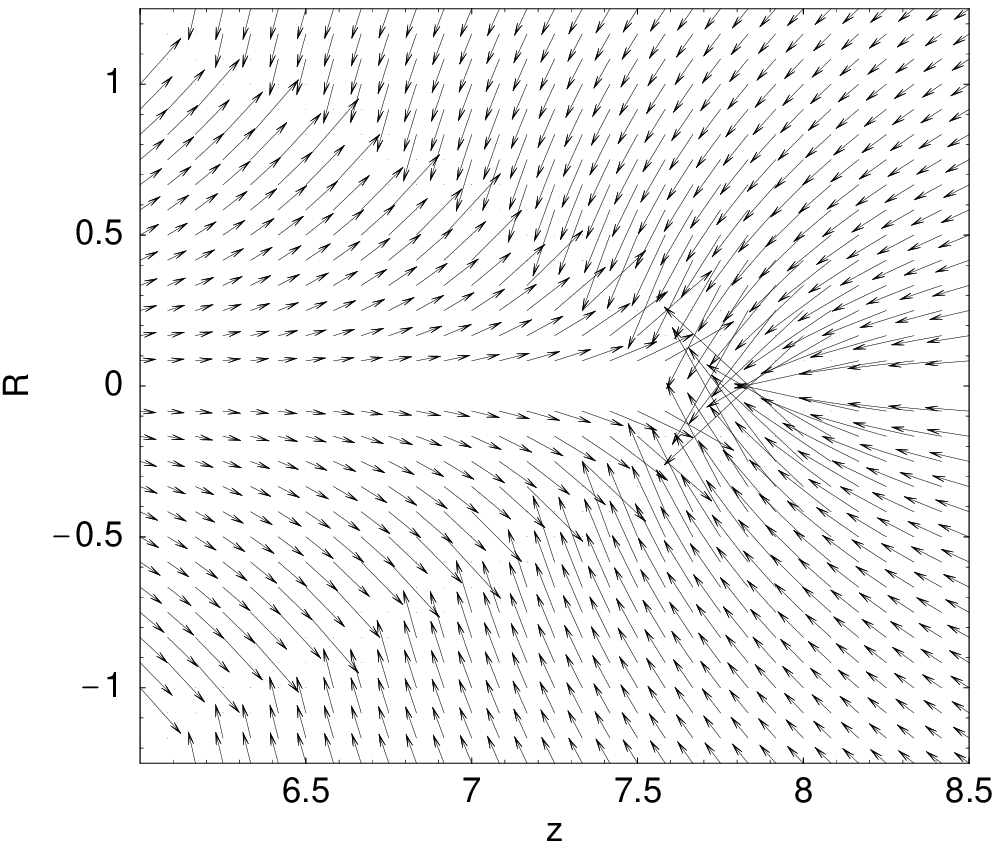}
\end{center}
\label{f8}
\caption{Gas flow induced by a perturber with uniform rectilinear motion is an infinite  gaseous medium. Distances are scaled to $V/c$ and velocities in each panel are magnified by a constant factor to show their relative magnitudes. Top left panel: the sonic wave front of a subsonic perturber with ${\cal M}=0.5$ travels as a shock in the ambient medium while the outside pressureless ballistic flow points toward $R=0$ and $z=Vt/2$ because of the symmetry of the corresponding potential. Top right panel: a zoom inside the sonic wave front around the perturber with ${\cal M}=0.5$. The velocity field is symmetric with respect to the trajectory and vanishes along it. Bottom left panel: the velocity field of a perturber that travel near the sound speed (${\cal M}=0.99$) is strongly peaked at the perturber's position. Bottom right panel: For a supersonic perturber (${\cal M}=2$), the velocity is maximal at the interface with the Mach cone.}
\end{figure}

\begin{figure}
\begin{center}
~\\[-12mm]
\includegraphics[width=75mm]{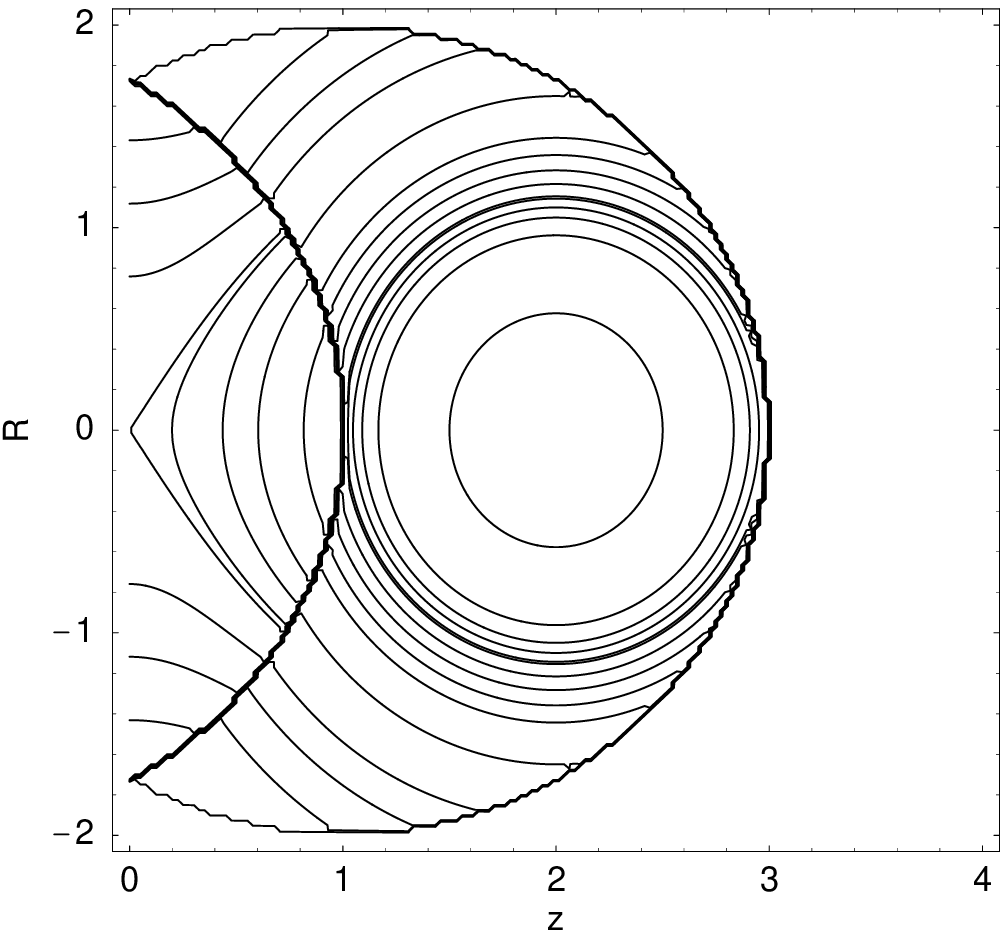}\includegraphics[width=75mm]{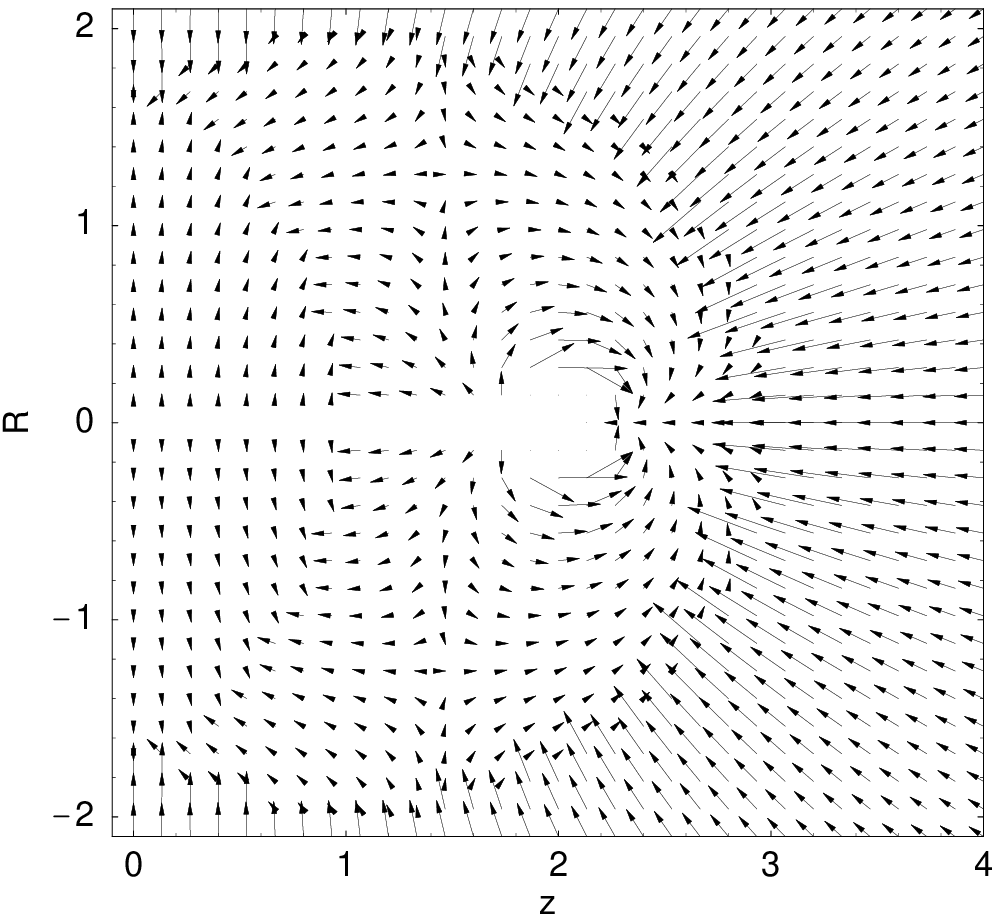}\\
\includegraphics[width=75mm]{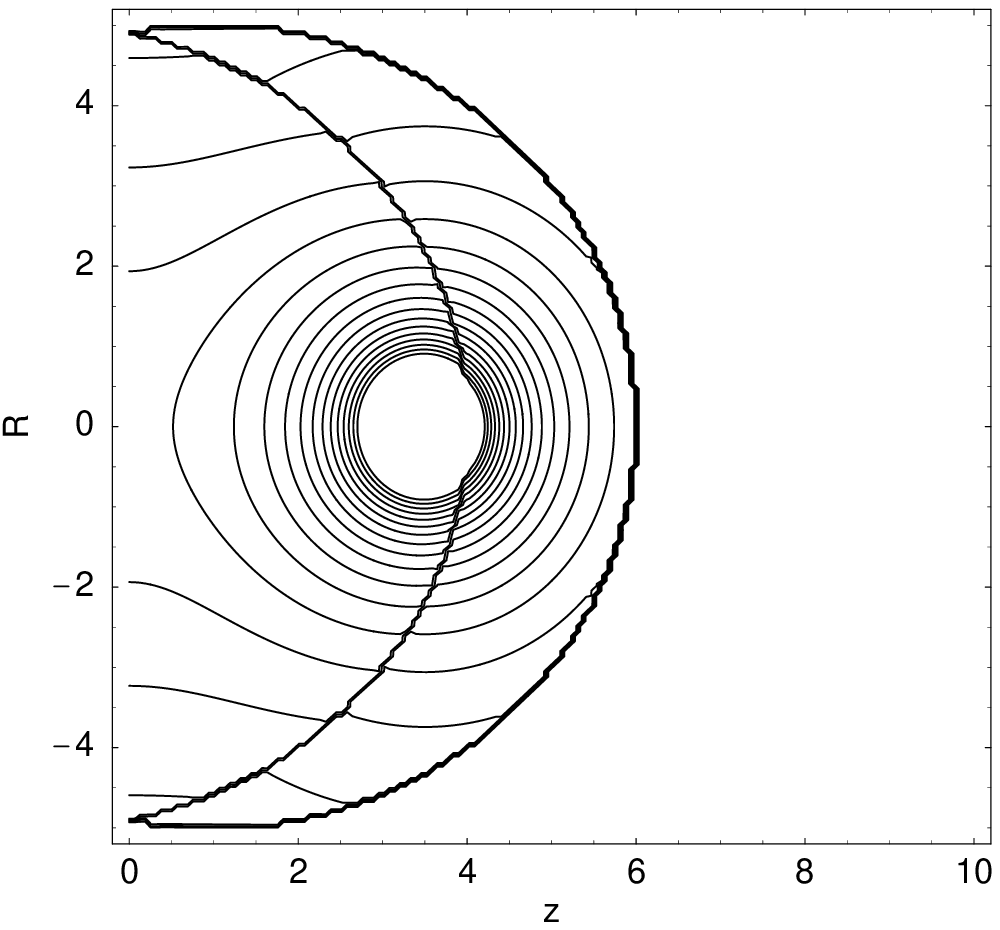}\includegraphics[width=75mm]{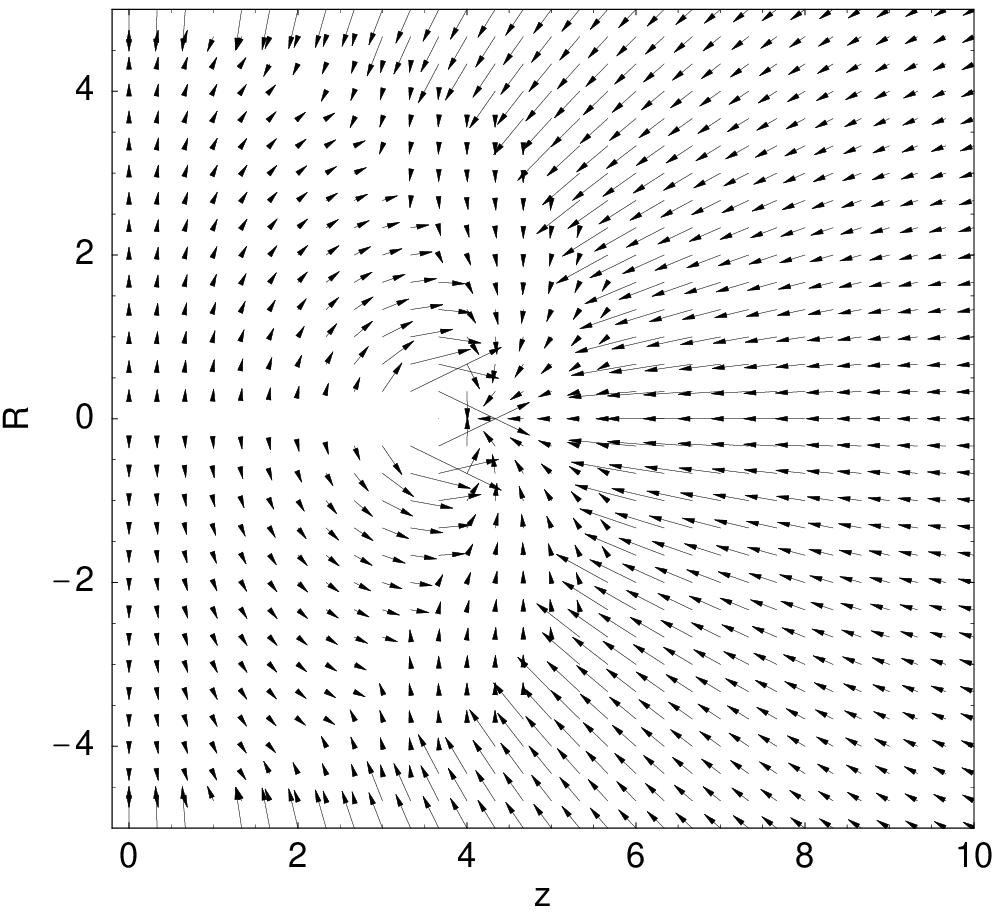}\\
\includegraphics[width=75mm]{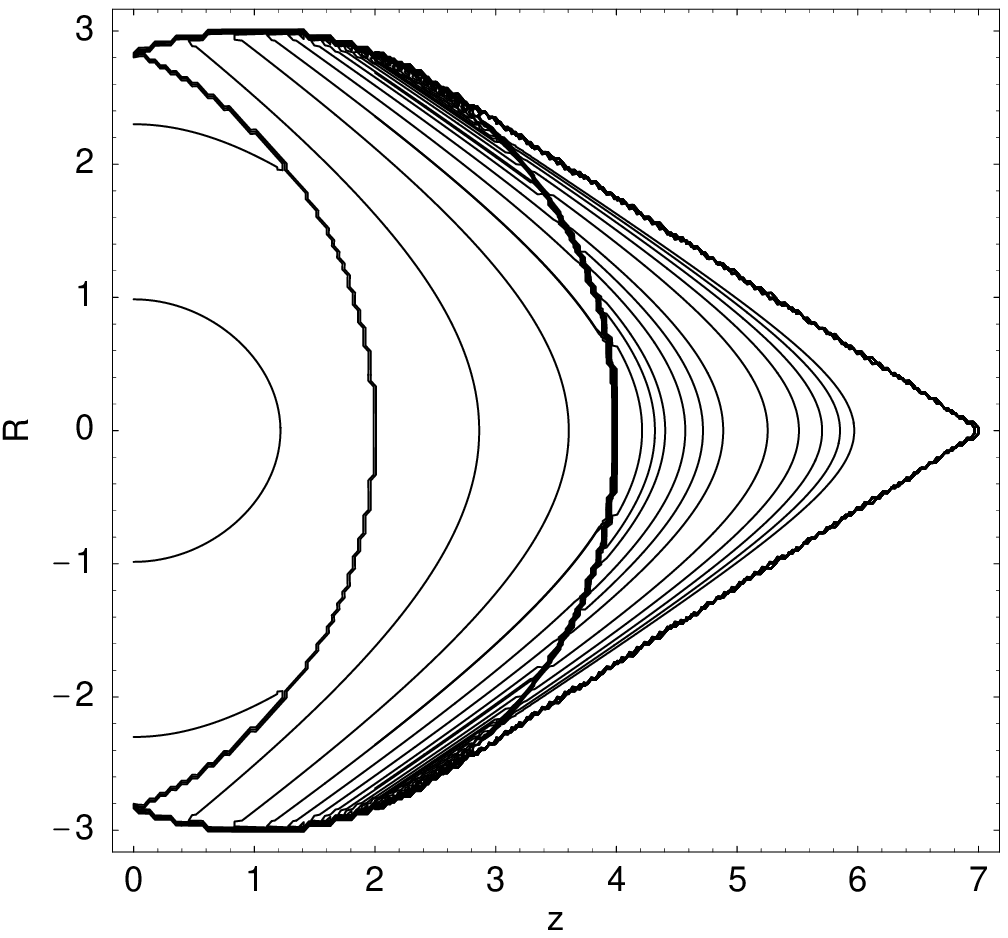}\includegraphics[width=75mm]{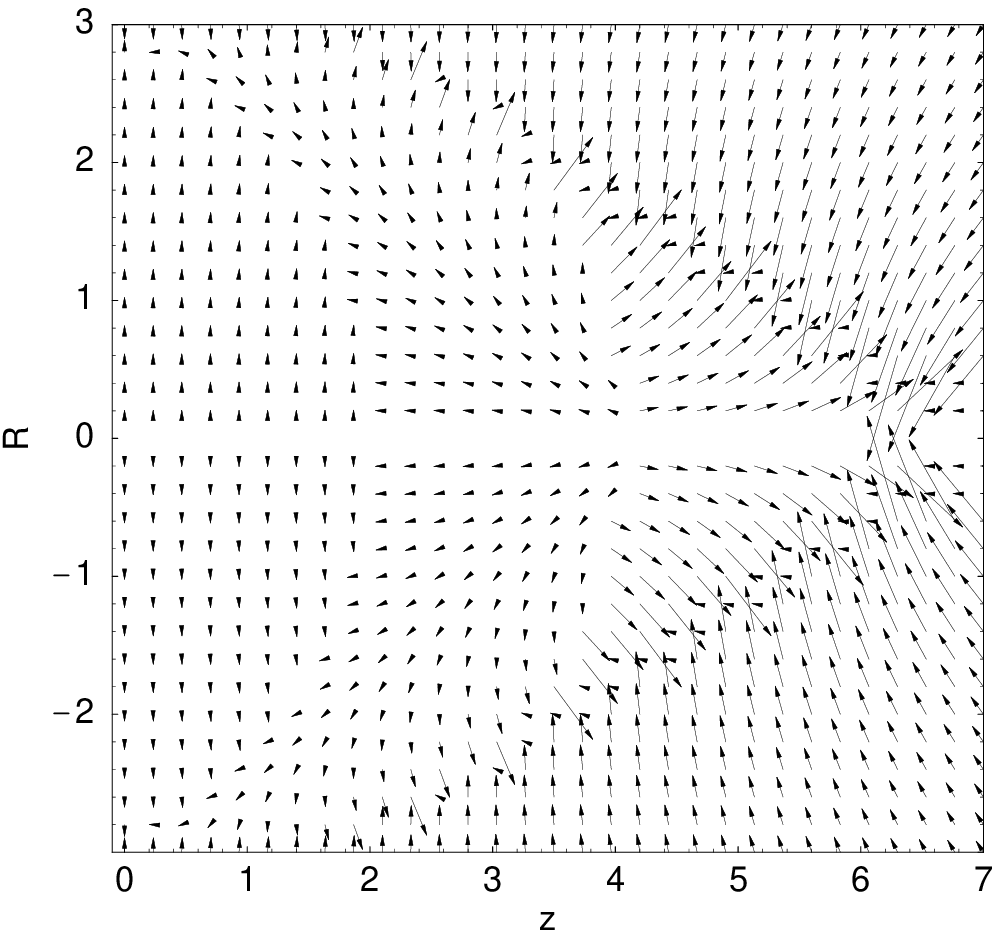}
\end{center}
\vspace*{-12mm}
\label{f9}
\caption{Density enhancement (left column) and velocity field (right column) for motion normal to the boundary at $z=0$. Distances are scaled to $d$ and time to $d/c$. Top row: A subsonic perturber with ${\cal M}=0.5$ at $t=2$; the reflected wake is behind the perturber. The solid lines denote the wavefront. Middle row: The reflected wake has overtaken the subsonic perturber ($t=5$). Bottom row: A supersonic perturber with ${\cal M}=2$.}
\end{figure}

\begin{figure}
\begin{center}
\includegraphics[width=75mm]{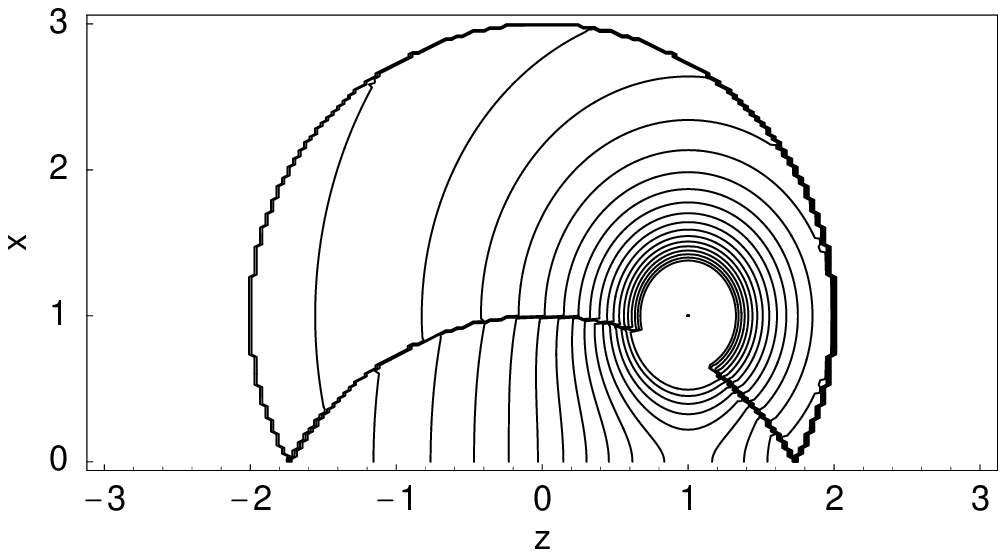}\includegraphics[width=75mm]{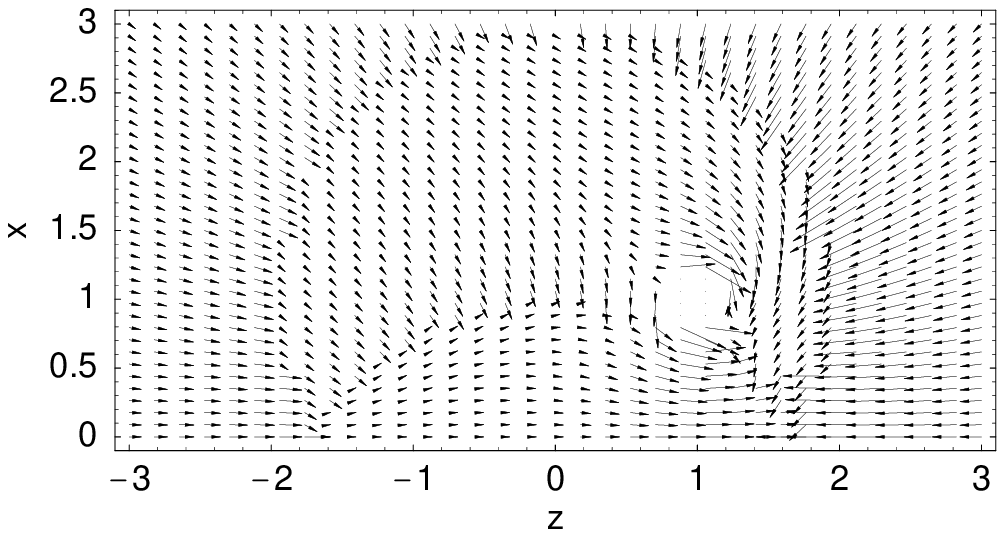}\\
\includegraphics[width=75mm]{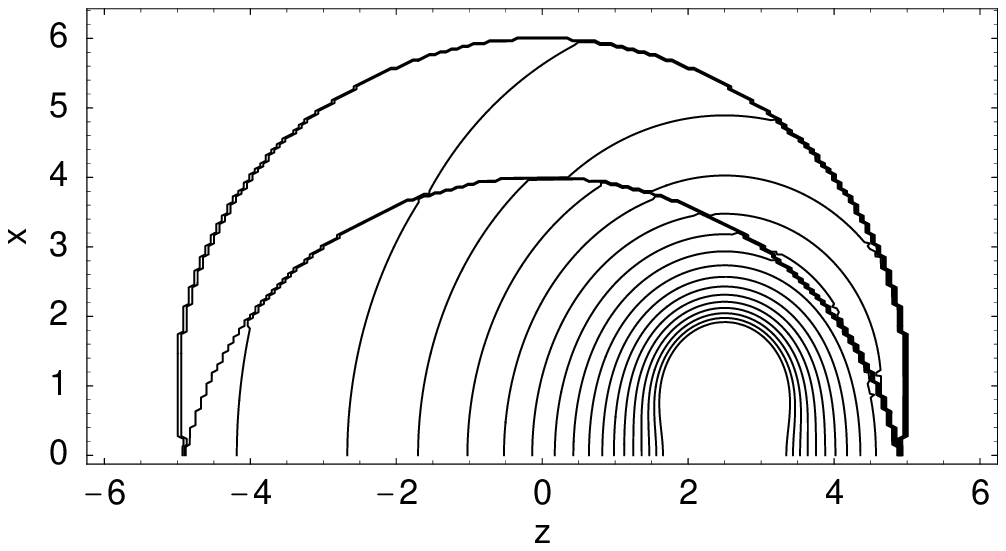}\includegraphics[width=75mm]{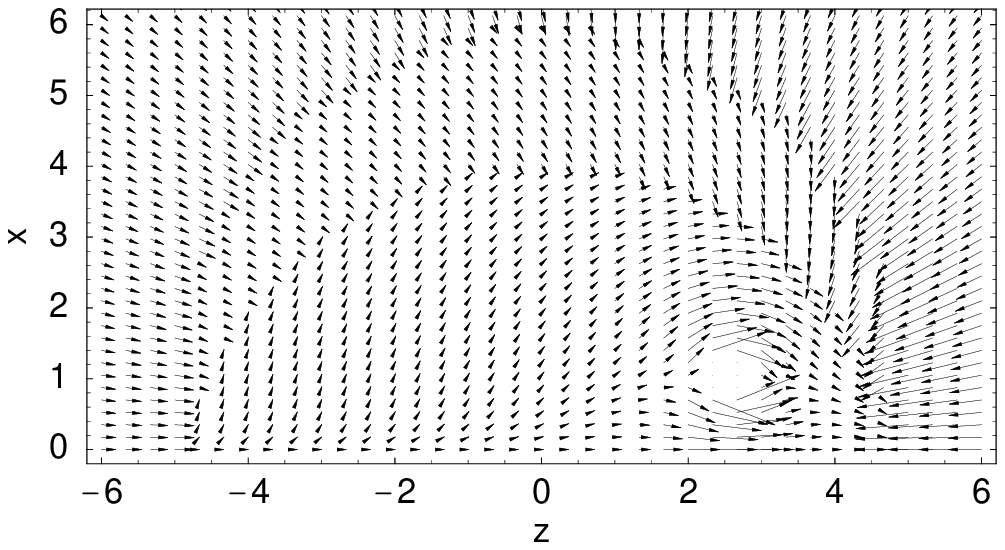}\\
\includegraphics[width=75mm]{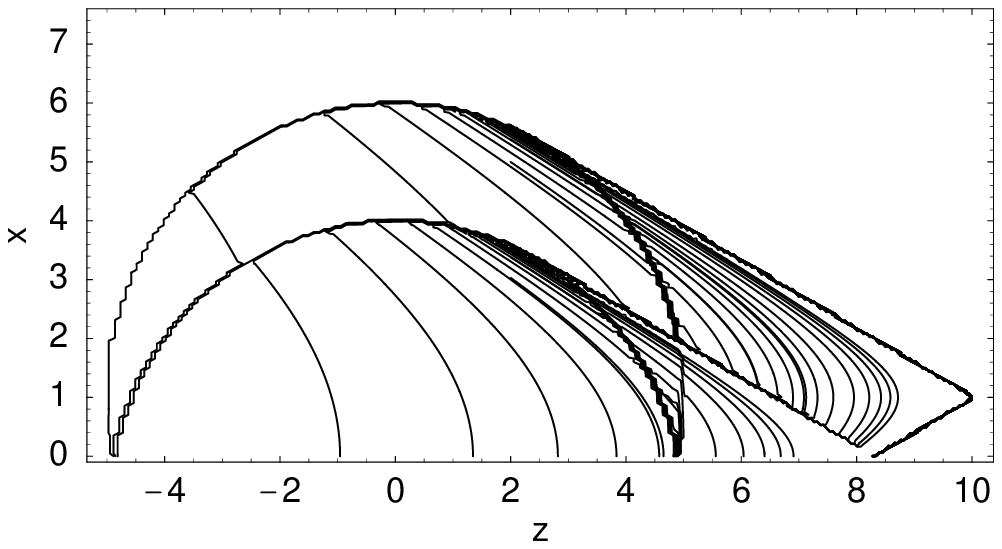}\includegraphics[width=75mm]{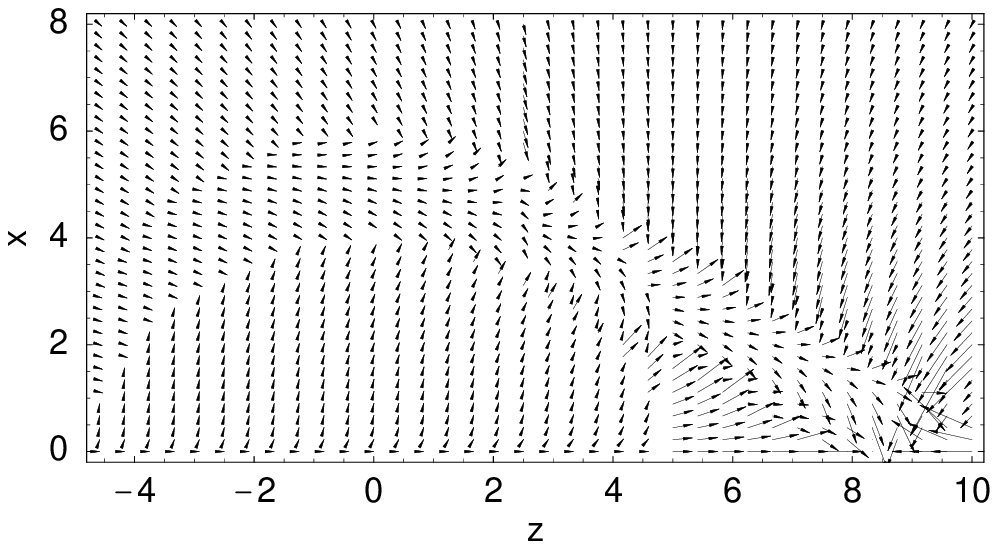}
\end{center}
\label{f10}
\caption{Density enhancement (left column) and velocity field (right column) for motion parallel to the boundary at $x=0$ in the plane $y=0$. Distances are scaled to $d$ and time to $d/c$. Top row: A subsonic perturber with ${\cal M}=0.5$ at $t=2$; the reflected wake is behind the perturber. The solid lines denote the wavefront. Middle row: The reflected wake has overtaken the subsonic perturber ($t=5$). Bottom row: A supersonic perturber with ${\cal M}=2$.}
\end{figure}


\begin{thebibliography}{}

\bibitem[\protect\citeauthoryear{Binney \& Tremaine}{2008}]{r5}
Binney J., Tremaine S., 2008, Galactic Dynamics, Princeton University Press, Princeton 

\bibitem[\protect\citeauthoryear{Bisnovatyi-Kogan et al.}{1979}]{r41} 
Bisnovatyi-Kogan G.~S., Kazhdan Y.~M.,  Klypin A.~A., Lutskii A.~E., Shakura N.~I., 1979, SvA, 23, 201 

\bibitem[Boylan-Kolchin et al.(2008)]{r43} Boylan-Kolchin, 
M., Ma, C.-P., \& Quataert, E.\ 2008, \mnras, 383, 93 

\bibitem[\protect\citeauthoryear{Bullock \& Johnston}{2005}]{r4} 
Bullock J.~S., Johnston K.~V., 2005, ApJ, 635, 931 

\bibitem[\protect\citeauthoryear{Chandrasekhar}{1943}]{r17} 
Chandrasekhar S., 1943, ApJ, 97, 255 

\bibitem[\protect\citeauthoryear{Cresswell et al.}{2007}]{r33} 
Cresswell P., Dirksen G., Kley W., Nelson R.~P., 2007, A\&A, 473, 329 

\bibitem[\protect\citeauthoryear{Dokuchaev}{1964}]{r21} 
Dokuchaev V.~P., 1964, SvA, 8, 23

\bibitem[\protect\citeauthoryear{Edgar}{2004}]{r37} 
Edgar R.~G., 2004, New Ast Rev 48, 843

\bibitem[\protect\citeauthoryear{El-Zant, Kim, 
\& Kamionkowski}{2004}]{r13} 
El-Zant A.~A., Kim W.-T., Kamionkowski M., 2004, MNRAS, 354, 169 

\bibitem[\protect\citeauthoryear{Faltenbacher et  al.}{2005}]{r14} 
Faltenbacher A., Kravtsov A.~V., Nagai D.,  Gottl{\"o}ber S., 2005, MNRAS, 358, 139 

\bibitem[\protect\citeauthoryear{Frenk et al.}{1996}]{r3} 
Frenk C.~S., Evrard A.~E., White S.~D.~M., Summers F.~J., 1996, ApJ, 472,  460 

\bibitem[\protect\citeauthoryear{Goldreich  \& Tremaine}{1982}]{r11} 
Goldreich P., Tremaine S., 1982, ARA\&A, 20, 249

\bibitem[\protect\citeauthoryear{Hahn}{2003}]{r30} 
Hahn  J.~M., 2003, ApJ, 595, 531 

\bibitem[\protect\citeauthoryear{H\'enon}{1958}]{r18} 
H\'enon M., 1958, AnAp, 21, 186

\bibitem[\protect\citeauthoryear{Horedt}{2000}]{r40} 
Horedt G.~P., 2000, ApJ, 541, 821 

\bibitem[\protect\citeauthoryear{Hornung, Pellat, \& Barge}{1985}]{r6} 
Hornung P., Pellat R., Barge P., 1985, Icar, 64, 295 

\bibitem[\protect\citeauthoryear{Ida}{1990}]{r8}
Ida S., 1990, Icar, 88, 129 

\bibitem[\protect\citeauthoryear{Just \& Kegel}{1990}]{r36} 
Just A., Kegel W.~H., 1990, A\&A, 232, 447 

\bibitem[\protect\citeauthoryear{Kim \& Morris}{2003}]{r1} 
Kim S.~S., Morris M., 2003, ApJ, 597, 312 

\bibitem[\protect\citeauthoryear{Kim, El-Zant, \& Kamionkowski}{2005}]{r15} 
Kim W.-T., El-Zant A.~A., Kamionkowski M., 2005, ApJ, 632, 157 

\bibitem[\protect\citeauthoryear{Kim \& Kim}{2007}]{r27} 
Kim H., Kim W.-T., 2007, ApJ, 665, 432 

\bibitem[\protect\citeauthoryear{Kim, Kim, \& S{\'a}nchez-Salcedo}{2008}]{r28} 
Kim H., Kim W.-T., S{\'a}nchez-Salcedo F.~J., 2008, ApJ, 679, L33 

\bibitem[\protect\citeauthoryear{Kokubo \& Ida}{1996}]{r9} 
Kokubo E., Ida S., 1996, Icar, 123, 180 

\bibitem[\protect\citeauthoryear{Larwood \& Papaloizou}{1997}]{r31} 
Larwood J.~D., Papaloizou J.~C.~B., 1997, MNRAS, 285, 288 


\bibitem[\protect\citeauthoryear{Namouni}{2010}]{r29} 
Namouni F., 2010, MNRAS, 401, 319 (Paper I)

\bibitem[\protect\citeauthoryear{Namouni, Luciani, \& Pellat}{1996}]{r10} 
Namouni F., Luciani J.~F., Pellat R., 1996, A\&A, 307, 972 

 \bibitem[\protect\citeauthoryear{Narayan}{2000}]{r12} 
Narayan R., 2000, ApJ, 536, 663 

\bibitem[\protect\citeauthoryear{Nelson et al.}{2000}]{r34} 
Nelson R.~P., Papaloizou J.~C.~B., Masset F., Kley W., 2000, MNRAS, 318, 18 

\bibitem[\protect\citeauthoryear{Ostriker \& Davidsen}{1968}]{r19} 
Ostriker J.~P., Davidsen A.~F., 1968, ApJ, 151, 679 

\bibitem[\protect\citeauthoryear{Ostriker}{1999}]{r24} 
Ostriker E.~C., 1999, ApJ, 513, 252 

\bibitem[\protect\citeauthoryear{Papaloizou, Nelson, \& Masset}{2001}]{r35} 
Papaloizou J.~C.~B., Nelson R.~P., Masset F., 2001, A\&A, 366, 263 

\bibitem[\protect\citeauthoryear{Portegies Zwart et al.}{2004}]{r2} 
Portegies Zwart S.~F., Baumgardt H., Hut  P., Makino J., McMillan S.~L.~W., 2004, Natur, 428, 724 

\bibitem[\protect\citeauthoryear{Rephaeli \& Salpeter}{1980}]{r23} 
Rephaeli Y., Salpeter E.~E., 1980, ApJ, 240, 20 

\bibitem[\protect\citeauthoryear{Ruderman \& Spiegel}{1971}]{r22} 
Ruderman M.~A., Spiegel E.~A., 1971, ApJ, 165, 1 

\bibitem[\protect\citeauthoryear{S{\'a}nchez-Salcedo \& Brandenburg}{1999}]{r25} 
S{\'a}nchez-Salcedo F.~J., Brandenburg A., 1999, ApJ, 522, L35 

\bibitem[\protect\citeauthoryear{S{\'a}nchez-Salcedo \& Brandenburg}{2001}]{r26} 
S{\'a}nchez-Salcedo F.~J., Brandenburg A., 2001, MNRAS, 322, 67 

\bibitem[\protect\citeauthoryear{S{\'a}nchez-Salcedo}{2009}]{r16} 
S{\'a}nchez-Salcedo F.~J., 2009, MNRAS, 392, 1573 

\bibitem[\protect\citeauthoryear{Stewart \& Wetherill}{1988}]{r7} 
Stewart G.~R., Wetherill G.~W., 1988, Icar, 74, 542 

\bibitem[Tanaka et al.(2002)]{r42} Tanaka, H., Takeuchi, 
T., \& Ward, W.~R.\ 2002, \apj \ 565, 1257 

\bibitem[\protect\citeauthoryear{Tremaine  \& Weinberg}{1984}]{r20} 
Tremaine S., Weinberg M.~D., 1984, MNRAS 209, 729

\bibitem[\protect\citeauthoryear{Ward}{2003}]{r32} 
Ward  W.~R., 2003, ApJ, 584, L39 

\bibitem[\protect\citeauthoryear{Wolfson}{1977}]{r38} 
Wolfson R., 1977, ApJ, 213, 208 

\bibitem[\protect\citeauthoryear{Yabushita}{1978}]{r39} 
Yabushita S., 1978, MNRAS, 183, 459 
\end{thebibliography}
\end{document}